\documentclass[a4paper,12pt]{article}

\usepackage{amsmath}
\usepackage{amssymb}
\usepackage{hyperref}
\numberwithin{equation}{section}
\numberwithin{figure}{section}
\numberwithin{table}{section}
\def\papertitlepage{\baselineskip 3.5ex\thispagestyle{empty}}
\def\Title#1{\baselineskip 1cm \vspace{1.5cm}%
  \begin{center}{\Large\bf #1}\end{center}\vspace{0.5cm}}
\def\Authors#1{\begin{center}\renewcommand{\thefootnote}{\fnsymbol{footnote}}{\it #1}\end{center}}
\def\Abstract{\vspace{1.0cm}%
  \begin{center}{\large\bf Abstract}\end{center}}
\renewenvironment{thebibliography}{\pagebreak[3]\par\vspace{0.6em}
\begin{flushleft}{\large \bf References}\end{flushleft}
\vspace{-1.0em}

\begin{enumerate}\if@twocolumn\baselineskip=0.6em\itemsep -0.2em
\else\itemsep -0.2em\fi\labelsep 0.1em}{\end{enumerate} }
\def\hbar{\Bar{h}}

\def\ybar{\Bar{y}}\def\zbar{\Bar{z}}

\def\calA{{\cal A}}

\def\calM{{\cal M}} \def\calN{{\cal N}}

\DeclareSymbolFont{rsfs}{U}{rsfs}{m}{n}
\DeclareSymbolFontAlphabet{\mathscr}{rsfs}

\def\del{\partial}
\def\mathd{\mathrm{d}}
\def\mathe{\mathrm{e}}
\def\mathi{i}

\def\Im{\mathop{\mathrm{Im}}\nolimits}

\newcommand{\fslash}[1]{\ooalign{\hfil\ensuremath{\slash}\hfil\crcr\ensuremath{#1}}}

\DeclareMathDelimiter{\lcolon}{\mathopen}{operators}{"3A}{largesymbols}{"3A}
\DeclareMathDelimiter{\rcolon}{\mathclose}{operators}{"3A}{largesymbols}{"3A}

\def\+{\!\!+\!\!}

\def\lpar{(\!(}
\def\rpar{)\!)}

\def\a{\alpha}
\def\t{\theta}

\def\bra<#1|{\langle#1|}
\def\ket|#1>{|#1\rangle}
\def\braket<#1|#2>{\langle#1|#2\rangle}
\def\llangle{\langle\!\langle}
\def\rrangle{\rangle\!\rangle}
\def\bbra<#1|{\llangle#1|}
\def\kket|#1>{|#1\rrangle}
\def\bbraket<#1|#2>{\llangle#1|#2\rrangle}
\begin{document}
{\papertitlepage
\hbox{ }\vspace*{0cm}
{
\hfill
\begin{minipage}{4.2cm}
IFT-P.002/2009\par\noindent
NSF-KITP-09-25\par\noindent
March, 2009
\end{minipage}}
\Title{Pure Spinor Vertex Operators in Siegel Gauge \\ and Loop
Amplitude Regularization}
\Authors{{\sc Yuri Aisaka${}^{1,2}$\footnote{\tt yuri@ift.unesp.br}}
and
{\sc Nathan~Berkovits${}^{1,2}$\footnote{\tt nberkovi@ift.unesp.br}}
\\
${}^1$Instituto de F\'{i}sica Te\'{o}rica,
S\~{a}o Paulo State University, \\[-2ex]
Rua Pamplona 145, S\~{a}o Paulo, SP 01405-900, Brasil
\\
${}^2$Kavli Institute for Theoretical Physics, Univ. of California at
Santa Barbara, \\[-2ex]
Santa Barbara, CA 93106-4030, USA
}

} 
\vskip-\baselineskip
\vskip-\baselineskip
{\baselineskip .5cm \Abstract

Since the $b$ ghost in the pure spinor formalism is a composite
operator depending on non-minimal variables, it is not trivial
to impose the Siegel gauge condition $b_0 V=0$ on BRST-invariant 
vertex operators. Using the antifield vertex operator $V^*$ 
of ghost-number $+2$, we show that Siegel gauge unintegrated vertex operators 
can be constructed as $b_0 V^*$ and Siegel gauge
integrated vertex operators as $\int dz ~b_{-1} b_0 V^*$.

These Siegel gauge vertex operators depend on the non-minimal variables,
so scattering amplitudes involving these operators need to be regularized
using the prescription developed previously with Nekrasov.
As an example of this regularization prescription, 
we compute the four-point one-loop amplitude with four Siegel gauge
integrated vertex operators. This is the first one-loop 
computation in the pure spinor formalism that does not require
unintegrated vertex operators. 
}

\newpage
\setcounter{footnote}{0}

\section{Introduction}

The pure spinor formalism~\cite{Berkovits:2000fe}
is a manifestly super-Poincar\'e covariant
description of the superstring which has been successfully used to
compute multiloop scattering amplitudes and covariantly quantize
Ramond-Ramond backgrounds. One of the most surprising features of
this formalism is that the $b$ ghost is not a fundamental worldsheet
variable but is a composite operator. Nevertheless, after replacing
the $b$ ghost with this composite operator,
the rules for computing scattering amplitudes are essentially the
same as in bosonic string theory.

In bosonic string theory, it is well-known that when a vertex
operator $V$ is in Siegel gauge, i.e. when $b_0 V=0$, 
the vertex operator is a conformal primary field. Since the scattering
amplitude prescription simplifies when the vertex operators are primary,
Siegel gauge is a convenient gauge choice. Furthermore, Siegel gauge is
a useful gauge choice in bosonic open string field theory since it
reduces the kinetic term $\langle \Phi Q \Phi \rangle$ to
$\langle \Phi c_0 L_0 \Phi \rangle$ so that the propagator is simply
${{b_0}\over{ L_0}}$.

Because the composite operator for the $b$ ghost depends on the
non-minimal variables in the pure spinor formalism, it is not immediately
obvious how to construct BRST-invariant vertex operators $V$ satisfying
the Siegel gauge condition $b_0 V=0$. For example, the massless open
string vertex operator in ``minimal'' gauge is $V=\lambda^\alpha
 A_\alpha(x,\theta)$, which does not satisfy $b_0 V=0$ for any choice
of $A_\alpha$. Note that $V=\lambda^\alpha A_\alpha(x,\theta)$
is a conformal primary whenever $\Box A_\alpha =0$ (which implies
Lorentz gauge for the gluon). So the condition $b_0 V=0$ in the
pure spinor formalism implies more than just the condition that $V$
is primary. 

As will be shown here, a natural way to construct vertex operators
in Siegel gauge is to start with the vertex operator $V^*$ for the antifield
which has ghost-number two. One then flips the statistics of the antifield
by defining $V^*$ to be a bosonic operator. Finally, one constructs the
unintegrated ghost-number one vertex operator $V_S$ in Siegel gauge as
$V_S = b_0 V^*$.
The corresponding integrated ghost-number zero vertex operator in
Siegel gauge is $\int dz~ b_{-1} b_0 V^*$.

This construction in bosonic string theory obviously reproduces
the usual Siegel gauge vertex operators where $V^* = c_0 V_S$.
But in the pure spinor formalism where there is no $c$ ghost, this
construction of Siegel gauge vertex operators is less trivial.
For example, since
the composite operator for $b$ depends on the non-minimal
variables, the resulting Siegel gauge vertex operator $V_S$ will
also depend on the non-minimal variables. 

The simplest example is the Siegel gauge massless vertex operator
which is constructed from the ghost-number two
operator $V^* = \lambda^\alpha \lambda^\beta 
A_{\alpha\beta}(x,\theta)$ where 
$A_{\alpha\beta}(x,\theta)$ is a bispinor superfield. As shown in~\cite{Berkovits:2001rb},
$QV^* =0$ and $\delta V^* = Q(\lambda^\alpha \Omega_\alpha)$ 
implies that the component fields
in 
$A_{\alpha\beta}(x,\theta)$ describe the antifields to the super-Yang-Mills
gluon and gluino. The corresponding Siegel gauge vertex operator
for the super-Yang-Mills multiplet is constructed in unintegrated form
as $b_0 V^*$, and in integrated form as $\int dz ~b_{-1} b_0 V^*$.

Since the composite operator for the $b$ ghost contains poles when 
the pure spinor variable satisfies $\lambda^\alpha=0$,
the vertex operator in Siegel gauge will also contain these poles.
As explained in~\cite{Berkovits:2004px}, these poles cause the functional integration over
$\lambda^\alpha$ to diverge if the order of the poles is greater than
or equal to 11. However, whenever this divergence occurs, the functional
integral over the fermionic non-minimal variables will vanish. The resulting
$0/0$ ambiguity can be regularized in a BRST-invariant manner using the
regularization prescription developed with Nikita Nekrasov in~\cite{Berkovits:2006vi}.

In this paper, we shall review how this regularization prescription works
for general multiloop scattering amplitudes. 
Furthermore, we will give the first non-trivial application
of this regularization procedure by computing a $4$-point one-loop amplitude
when all four vertex operators are chosen in Siegel gauge in integrated form.
Note that, as in bosonic string theory, $N$-point one-loop 
amplitudes can be computed using $N$ integrated vertex operators only if
all of the vertex operators are in Siegel gauge. So all previous one-loop
computations using the pure spinor formalism required at least one
unintegrated vertex operator. 

It is possible that this new one-loop 
amplitude prescription  
will be useful for comparing with the operator approach in the
pure spinor formalism, or
with other superstring prescriptions such as
the Lee-Siegel~\cite{Lee:2005jy} or RNS prescriptions.
Another possible application of our results is for 
super-Poincar\'e covariant
open superstring field theory. Although 
a cubic
open superstring field theory action has
been constructed using the pure spinor formalism~\cite{Berkovits:2005bt},
the gauge-fixing of this action has not yet been performed. It
seems likely that the gauge-fixing techniques developed here will also
be useful for gauge-fixing the open superstring field theory action.

\bigskip
This paper is organized as follows.
In section~\ref{sec:review},
the basics of the pure spinor formalism is reviewed
and, in section~\ref{sec:regularization},
the regularization method proposed in~\cite{Berkovits:2006vi}
to deal with $(\overline{\lambda}\lambda)$ poles is explained.
Section~\ref{sec:siegel}
deals with the construction of the vertex operators
in the Siegel gauge.
Both unintegrated and integrated vertex operators
are described, and some of their properties are studied.
In section~\ref{sec:application},
the Siegel gauge vertex operators are used to
define a new $n$-point $1$-loop amplitude prescription
that uses only integrated vertex operators.
In addition,
the regularization
of~\cite{Berkovits:2006vi} is explained using
the example of the $4$-point $1$-loop amplitude.
We conclude in section~\ref{sec:summary}
and indicate some directions for future works.

\bigskip
{\bf Note:} While this paper was being written up, we received a draft
of a paper by P.A. Grassi and P. Vanhove which also discusses
Siegel gauge vertex operators and regularization in the pure spinor formalism.
However, their discussion differs considerably from our paper.
At the end of section 3, we have added some comments related to
their paper \cite{grassi:2009gv}
which appeared shortly after the original version of
this paper.

\section{A review of the pure spinor formalism}
\label{sec:review}

We begin by reviewing certain aspects of the
pure spinor formalism that are relevant to the present paper.

\subsection{World sheet fields}

Field contents of the worldsheet theory of the pure spinor formalism
can be divided into matter and ghost sectors.
The former consists
of the Green-Schwarz-Siegel variables~\cite{Siegel:1985xj}
\begin{align}
(x^m;\, p_{\alpha},\,\theta^{\alpha}),\quad
 m=0,\cdots,9\,,\quad \alpha=1,\cdots,16
\end{align}
that describe the embedding of the string
in a superspace $(x^{m},\,\theta^{\alpha})$.\footnote{%
Throughout, we shall use the notation appropriate
for describing a chiral half of the closed string theory,
but use a terminology appropriate for the open string.}
They satisfy free field OPEs
\begin{align}
x^{m}(z)x^{n}(w) &= -\eta^{mn}\log(z-w)\,,\quad
p_{\alpha}(z)\theta^{\beta}(w) = {\delta_{\alpha}{}^{\beta}\over z-w}\,.
\end{align}

The ghost sector consists of a conjugate pair of bosonic spinors
\begin{align}
(\omega_{\alpha}, \lambda^{\alpha})\,,
\end{align}
but they must be treated with care
as they are not genuine free fields;
instead, $\lambda^{\alpha}$ (weight $0$)
is subject to the so-called pure spinor constraint
\begin{align}
\label{eq:PSconstraint}
\lambda^{\alpha}\gamma^{m}_{\alpha\beta} \lambda^{\beta} &= 0\,,
\end{align}
where $\gamma^{m}_{\alpha}$ is the symmetric $16\times16$
gamma matrices in ten dimensions.

To be consistent with this constraint,
the conjugate $\omega_{\alpha}$ (weight $1$) is
defined up to a gauge transformation
\begin{align}
\label{eq:PSgaugeinv}
\delta \omega_{\alpha}(z) &= \Omega_{m}(z)(\gamma^{m}\lambda)_{\alpha}\,,
\end{align}
and $\omega_{\alpha}$ can only appear in gauge invariant combinations.
Some basic invariants of this gauge transformation
are $\lambda$-charge current $J_{\lambda}$, Lorentz generator $N_{mn}$
and energy-momentum tensor $T_{\lambda}$
defined as
\begin{align}
J_{\lambda} &= \omega\lambda\,,\quad N_{mn} = {1\over2}(\omega\gamma_{mn}\lambda)\,,\quad
T_{\lambda} = \omega\del \lambda\,.
\end{align}

The OPE algebra formed by those basic gauge invariants
can be computed by parameterizing the components
of $\lambda^{\alpha}$ and $\omega_{\alpha}$ by $U(5)$ covariant genuine free fields.
The resulting algebra is
\begin{align}
\begin{split}
N^{mn}(z)\lambda^{\alpha}(w) &= {{1\over2}(\gamma^{mn}\lambda)^{\alpha}(w) \over z-w}\,,\quad
J_{\lambda}(z)\lambda^{\alpha}(w) = {\lambda^{\alpha}(w)\over z-w}\,, \\
N^{mn}(z)N^{pq}(w)
 &= {-3(\eta^{np}\gamma^{mq}-\eta^{mp}\eta^{nq})\over (z-w)^2}
 + {\eta^{np}N^{mq}(w)+(\text{$3$-terms})\over z-w}\,, \\
J_{\lambda}(z)J_{\lambda}(w) &= {-4\over(z-w)^{2}}\,,\quad
J_{\lambda}(z)N^{mn}(w) = \text{regular}\,, \\
N_{mn}(z)T_{\lambda}(w) &= {N_{mn}(w)\over (z-w)}\,,\quad
J_{\lambda}(z)T_{\lambda}(w) = {-8\over(z-w)^{3}}+{J_{\lambda}(w)\over (z-w)^{2}}\,, \\
T_{\lambda}(z)T_{\lambda}(w)
&= {11\over(z-w)^{4}}+{2T_{\lambda}(w)\over(z-w)^{2}}
  + {\del T_{\lambda}(w)\over z-w} \,.
\end{split}
\end{align}

\subsection{Physical states}

Physical open string states
are defined as the ghost number ($\lambda$-charge)
$1$ cohomology of a nilpotent BRST operator
\begin{align}
Q &= \oint \lambda^{\alpha}d_{\alpha}
\end{align}
where
\begin{align}
d_{\alpha}=p_{\alpha}+{1\over2}(\gamma_{m}\theta)_{\alpha}\del x^{m}
 - {1\over8}(\gamma_{m}\theta)_{\alpha}(\theta\gamma^{m}\del\theta)
\end{align}
has the form of the phase space constraint of
the classical Green-Schwarz action.
Using the free field OPE between $p_{\alpha}$ and $\theta^{\alpha}$,
$d_{\alpha}$ satisfies
\begin{align}
d_{\alpha}(z)d_{\beta}(w) &= {\Pi_{m}\gamma^{m}_{\alpha\beta}\over z-w}
\end{align}
where $\Pi_{m}=\del x_{m} + {1\over2}(\theta\gamma^{m}\del\theta)$
is the supersymmetric momentum.
$d_{\alpha}$ and $\Pi_{m}$ acts on superfields
as supercovariant derivatives:
\begin{align}
D_{\alpha} &= \del_{\alpha} - {1\over2}(\gamma^{m}\theta)_{\alpha}\del_{m},\quad
P_{m} = -\del_{m}\,.
\end{align}

For example, massless states are described by the
$\lambda$-charge $1$ vertex operator
\begin{align}
V &= \lambda^{\alpha}A_{\alpha}(x,\theta)\,.
\end{align}
The cohomology condition $QV=0$ and $\delta V=Q\Omega$
implies that the superfield $A_{\alpha}$ satisfies
the correct on-shell constraint $\gamma^{\alpha\beta}_{m_{1}\cdots m_{5}}D_{\alpha}A_{\beta}=0$
and gauge invariance $\delta A_{\alpha}=D_{\alpha}\Omega$.

Although the form of the BRST operator above appears strange
at first sight, its ghost number $1$ cohomology can be
explicitly studied using an $SO(8)$ parameterization
of pure spinors and it reproduces the lightcone spectrum
of the Green-Schwarz superstring~\cite{Berkovits:2000nn}.
Moreover, there are arguments that it can in fact be derived
from the classical
Green-Schwarz action~\cite{Tonin:2002tt,Berkovits:2004tw,Aisaka:2005vn,
garcia:2005gg,Berkovits:2007wz},
and that it is related to the BRST operators of
Ramond-Neveu-Schwarz and Green-Schwarz formalisms
by similarity transformations~\cite{Aisaka:2002sd}.

Finally, $Q$ has cohomologies at other $\lambda$-charges as well.
They are interpreted as spacetime ghosts, antifields
and antighosts.

\subsection{Pure spinor sector as a curved $\beta\gamma$ system: non-minimal formalism}

At first sight, handling a non-linearly constrained system
such as the pure spinor system appears difficult.
However, it can be treated rigorously using the
theory of curved $\beta\gamma$-systems~\cite{Malikov:1998dw}
(if the origin $\lambda\equiv 0$ of the pure spinor space
is removed~\cite{Nekrasov:2005wg}).
One way to apply this idea to the pure spinor formalism
is to introduce another set of pure spinors
and its fermionic partners,
$(\overline{\omega}^{\alpha},\overline{\lambda}_{\alpha};s^{\alpha},r_{\alpha})$.

$\overline{\lambda}_{\alpha}$ is an antichiral pure spinor (weight $0$)
\begin{align}
\overline{\lambda}_{\alpha}\gamma_{m}^{\alpha\beta}\overline{\lambda}_{\beta} = 0\,,
\end{align}
$r_{\alpha}$ is a fermionic field (weight $0$) that is constrained as
\begin{align}
 r_{\alpha}\gamma_{m}^{\alpha\beta}\overline{\lambda}_{\beta} =0\,,
\end{align}
and $\overline{\omega}^{\alpha}$ and $s^{\alpha}$ are
the conjugate momenta (weight $1$)
of $\overline{\lambda}_{\alpha}$ and $r_{\alpha}$, respectively.
In ten dimensional Euclidean space,
$\overline{\lambda}_{\alpha}$ can be regarded as the complex conjugate of $\lambda^{\alpha}$,
and $r_{\alpha}$ is an extension of the target space differential
of $\overline{\lambda}_{\alpha}$:
\begin{align}
r_{\alpha} \sim \mathd\overline{\lambda}_{\alpha}\,.
\end{align}

Just as $\omega_{\alpha}$ must appear in
invariant combinations under
the gauge transformation $\delta\omega_{\alpha}=\Omega_{m}(\gamma^{m}\lambda)_{\alpha}$,
the conjugates $\overline{\omega}^{\alpha}$ and $s^{\alpha}$ must
appear in invariant combinations under
\begin{align}
\delta\overline{\omega}^{\alpha}
&= \overline{\Omega}_{m}(\gamma^{m}\overline{\lambda})^{\alpha}
 - \phi_{m}(\gamma^{m}r)^{\alpha}\,,\quad
\delta s^{\alpha}
= \phi_{m}(\gamma^{m}\overline{\lambda})^{\alpha}\,,
\end{align}
for arbitrary $\overline{\Omega}_{m}$ and $\phi_{m}$.
Some basic invariants are
\begin{align}
\label{eqn:NMcurs}
\begin{split}
\overline{N}_{mn} &= {1\over2}(\overline{\omega}\gamma_{mn}\lambda-s\gamma_{mn}r)\,,\quad
\overline{J}_{\overline{\lambda}} = \overline{\omega}\overline{\lambda} - sr\,,\quad
T_{\overline{\lambda}} = \overline{\omega}\del\overline{\lambda} - s\del r\,,
\\
J_{r}& = -sr\,,\quad S_{mn} = {1\over2}(s\gamma_{mn}\overline{\lambda})\,,\quad
 S = s\overline{\lambda}\,.
\end{split}
\end{align}

By parameterizing non-minimal variables
by $U(5)$ covariant free fields
(antiholomorphic local coordinates on the pure spinor space
and their conjugates),
OPE's among the basic invariants can be computed.
In particular, they satisfy
\begin{align}
\begin{split}
J_{r}(z)J_{r}(w) &= {11\over(z-w)^{2}}\,,\quad
\overline{J}_{\overline{\lambda}}(z)J_{r}(w) = {8\over(z-w)^{2}}\,,\quad
\overline{J}_{\overline{\lambda}}(z)\overline{J}_{\overline{\lambda}}(w)=\text{regular}\,,
\\
J_{r}(z)T_{\overline{\lambda}}(w)
 &= {11\over(z-w)^{3}} + {J_{r}(w)\over (z-w)^{2}}\,,\quad
\overline{J}_{\overline{\lambda}}(z)T_{\overline{\lambda}}(w)
 = {\overline{J}_{\overline{\lambda}} \over (z-w)^{2}}\,,
\\
T_{\overline{\lambda}}(z)T_{\overline{\lambda}}(w)
&= {2T_{\overline{\lambda}}(w) \over (z-w)^{2}}
 + {\del T_{\overline{\lambda}}(w) \over z-w} \,.
\end{split}
\end{align}
Therefore, the addition of the non-minimal sector does
not affect the total central charge,
but the total ghost number anomaly is shifted to
$3=11-8$, if one defines the ghost number by
\begin{align}
J_{g} = J_{\lambda} - J_{r} -{\overline J}_{\overline{\lambda}}
= \omega\lambda-\overline\omega\overline\lambda\,.
\end{align}

Physical states are then redefined as
the ghost number $1$ cohomology of
a nilpotent BRST operator
\begin{align}
\begin{split}
Q &= Q_{0} + Q_{1} \\
Q_{0} &= \oint\lambda^{\alpha}d_{\alpha}\,,\quad
Q_{1} = \oint r_{\alpha}\overline{\omega}^{\alpha} \,.
\end{split}
\end{align}
Subscripts denote the $r$-charge,
but note that both $Q_{0}$ and $Q_{1}$
carry $+1$ charge
under the ghost number current $J_{g}=\omega\lambda-
\overline\omega\overline\lambda$.

The additional piece $Q_{1}$
of the BRST operator deals with the constrained nature
of the pure spinor, and is essential for
having a composite $b$-ghost operator that satisfies
\begin{align}
T &= \{ Q\,, b\}\,,
\end{align}
for the total energy-momentum tensor $T$.
For future reference, we here record the explicit form of the
$b$-ghost:
\begin{align}
\begin{split}
b &= b_{-1} + b_0 + b_1 + b_2 + b_3 \\
b_{-1} &= s^{\alpha}\del\overline{\lambda}_{\alpha} \\
b_0 &= { \overline{\lambda}_{\alpha}
  [2\Pi^{m}(\gamma_{m}d)^{\alpha}
  - N_{mn}(\gamma^{mn}\del\theta)^{\alpha}
  - J_{\lambda}\del\theta^{\alpha}
 - {1\over4}\del^2\theta^{\alpha}]
\over 4(\overline{\lambda}\lambda)} \\
b_1
&= {(\overline{\lambda}\gamma^{mnp} r)
    (d\gamma_{mnp} d +24 N_{mn}\Pi_p)
 \over{192(\overline{\lambda}\lambda)^2}} \\
b_2
&= { (r\gamma_{mnp} r)(\overline{\lambda}\gamma^m d)N^{np}
  \over 16(\overline{\lambda}\lambda)^3} \\
b_3
&= {(\overline{\lambda}\gamma^{mnp}r)(r\gamma_{p}{}^{qr}r) N_{mn} N_{qr}
  \over 128(\overline{\lambda}\lambda)^4}\,.
\end{split}
\end{align}

Now, $Q_{1}$ itself is nilpotent and its cohomology
can be regarded as the operator space of the pure spinor
sector.
In~\cite{Aisaka:2008vw,Aisaka:2008vx,AldoArroyo:2008zm}, the structure of this
operator space was studied by computing its partition function.
One outcome of the investigation was that
$Q_{1}$-cohomology consists of two sectors
$H^{0}(Q_{1})$ and $H^{3}(Q_{1})$,
and there is a one-to-one mapping between the two.
(Here, the degree of the cohomology is the differential
form degree carried by $r_{\alpha}\sim\mathd\overline{\lambda}_{\alpha}$.)
An important element of $H^{3}(Q_{1})$ is the
tail term $b_{3}$ of the composite $b$-ghost.
It was found that $H^{3}(Q_{1})$ is essential for having
the operator doubling between spacetime fields and antifields,
and found that the total cohomology $H^{\ast}(Q_{1})$
has precisely the right structure
to kill the unphysical degrees of freedom contained
in the covariant oscillators $(x^{m},p_{\alpha},\theta^{\alpha})$,
up to the fifth mass level.

\bigskip
Having introduced the basic ingredients of the pure spinor
formalism, we now turn to the description of
scattering amplitudes.

\subsection{Tree amplitude}

To compute $n$-point tree amplitudes,
one uses $3$ unintegrated vertex operators $V$
and $n-3$ integrated vertex operators $\int\mathd z U(z)$
as in the bosonic string,
where $U$ carries weight $1$, ghost number $0$,
and is related to $V$ as $QU=\del V$:
\begin{align}
\calA_{n} = \int \prod_{i=4}^{n}\mathd^{2}w_{i}
|\langle V_{1}(z_{1})V_{2}(z_{2})V_{3}(z_{3})
 \prod_{i=4}^{n}U(w_{i})\rangle|^{2}
\end{align}
Here, $\langle\cdots\rangle$
denotes functional integrations over
$(\theta^{\alpha},\lambda^{\alpha},\overline{\lambda}_{\alpha},r_{\alpha})$.
(We ignore the functional integration over $x^{m}$
for simplicity because there is nothing special about it
in the pure spinor formalism.)

After integrating out the non-zero modes
using OPE's,
one is left with the zero-mode integration of the form
\begin{align}
\label{eqn:leftone}
\calA &= \int[\mathd \lambda][\mathd\overline{\lambda}][\mathd r]\mathd^{16}\theta
~ \calN f(\lambda,\overline{\lambda},r,\theta)\,,
\end{align}
where $f(\lambda,\overline{\lambda},r,\theta)$ is a function of the zero modes,
and the zero mode measures behave as
\begin{align}
[D\lambda] & = \lambda^{-3}\mathd^{11}\lambda\,,\quad
[D\overline{\lambda}] = \overline{\lambda}^{-3}\mathd^{11}\overline{\lambda}\,,\quad
[D r] = \overline{\lambda}^{3}\mathd^{11}r \,.
\end{align}

Now, this zero mode integration is ambiguous
due to an indefinite factor $\infty\cdot0$
that comes from non-compact bosonic integration
over $\lambda^{\alpha}$ and $\overline{\lambda}_{\alpha}$,
and unsaturated fermionic integration over $\theta^{\alpha}$ and $r_{\alpha}$.
However, this difficulty can be easily overcome
by inserting a regularization factor of the form
\begin{align}
\calN_{0} = \exp\{Q,\chi\}\,.
\end{align}
Since $\calN_{0} = 1 + Q\Omega$ for some $\Omega$,
$\calN_{0}$ respects BRST symmetry,
and one is free to choose whatever $\chi$ that is convenient for
computing amplitudes.
A simple and convenient choice is
\begin{align}
\chi &= -\overline{\lambda}\theta\,,
\end{align}
which leads to
\begin{align}
\calN_{0} = \exp(-\overline{\lambda}\lambda-r\theta)\,.
\end{align}
$\calN_{0}$ puts an exponential cut-off for bosonic zero-modes,
and provides extra fermionic zero-modes via
an expansion of the exponential.
By now, it is well tested that
this prescription lead to correct 
tree amplitudes
(see for example~\cite{Mafra:2009wq} for a review).

\subsection{Loop amplitudes}

We now turn to the discussion of loop amplitudes.
A natural prescription to use for the
$n$-point $g$-loop amplitude is
\begin{align}
\calA_{g}
&= \int_{\calM_{g,n}}\mathd^{3g-3}\tau_{k}
  \prod_{i=1}^{n}\int\mathd^{2}w_{i}
  \prod_{k=1}^{3g-3}\int\mathd^{2}z_{k}
 |\langle\prod_{k=1}^{3g-3}(\mu_{k}\cdot b)(z_{k})\prod_{i=1}^{n}U(w_{i})
 \rangle|^{2} \,,
\end{align}
where $\tau_{k}$ and $\mu_{k}$ are the Teichm\"uller parameters
and associated Beltrami operator,
and the bracket $\langle\cdots\rangle$
denotes functional integrations over the worldsheet fields.
However, this prescription is incomplete
as there are two subtleties
with the functional integration over zero modes.

The first subtlety comes from the proper definition
of an $\infty\cdot0$ factor associated with
the integration over non-compact bosonic zero modes,
and over unsaturated fermionic zero modes.
This indefinite factor can be defined as before
by inserting an operator of the form $\calN_{0}=\exp\{Q,\chi_{0}\}$
to the zero mode integral.
The only modification needed is that now
there are zero modes for weight $1$ operators
$(N_{mn},\overline{N}_{mn},J_{\lambda},\overline{J}_{\overline{\lambda}},S_{mn},S)$
as well.

A convenient choice for $\chi_{0}$ is
\begin{align}
\chi_{0} &= -\overline{\lambda}\theta - {1\over2}\sum_{I=1}^{g}(N_{I,mn}S_{I}^{mn}+J_{I} S_{I})
\end{align}
where $(N_{I,mn},J_{I};S_{I,mn},S_{I})_{I=1\sim g}$
are the $g$ zero modes of the corresponding operators
defined in~(\ref{eqn:NMcurs}).
For this choice, the zero mode integration comes
with an insertion of
\begin{align}
\begin{split}
\calN_{0}
&= \exp[-(\overline{\lambda}\lambda + r\theta)] \\
&\qquad\times
  \exp\sum_{I=1}^{g}[- {1\over2}N_{I,mn}\overline{N}^{mn}_{I}
  - J_{I}\overline{J}_{I}
   - {1\over2}S_{I,mn}(d_{I}\gamma^{mn}\lambda)
   - S_{I}(\lambda d_{I})] \,.
\end{split}
\end{align}
$\calN_{0}$ puts an exponential cut-off for bosonic zero-modes,
and provides extra fermionic zero-modes via
an expansion of the exponential.

\bigskip
There is another (possible) source of an indefinite factor $\infty\cdot0$
coming from the integration around $(\overline{\lambda}\lambda)\sim 0$.
This second subtlety is due to the $(\overline{\lambda}\lambda)$ poles
in the integrand
that could come from the measure,
from the insertion of the composite $b$-ghosts,
and from the vertex operators.
When the order of the $(\overline{\lambda}\lambda)$ pole is too high,
one gets a divergence upon integrating
near $(\overline{\lambda}\lambda)\sim0$.
However, this type of divergence always
comes with a zero coming from an over saturation
of $r$ zero modes,
and it was shown in~\cite{Berkovits:2006vi} how to regularize
and define this indefinite factor.

Since this regularization for $(\overline{\lambda}\lambda)\sim0$
is slightly involved, we shall explain it in
a separate (next) section.
In the absence of the second subtlety
at $(\overline{\lambda}\lambda)\to0$,
the prescription above is well tested
and known to lead to correct loop amplitudes~\cite{Berkovits:2005df}.
(See~\cite{Mafra:2009wq} for a review on the subject,
and~\cite{Anguelova:2004pg} for an extension of loop computations
to eleven dimensions.)

\section{Regularization of $(\overline{\lambda}\lambda)\sim0$}
\label{sec:regularization}

In this section, we explain the regularization prescription
of~\cite{Berkovits:2006vi} for the functional integral
region $(\overline{\lambda}\lambda)\sim 0$.
Although the basic idea of~\cite{Berkovits:2006vi}
is simple, the formulas there ended up complicated
because one had to make the prescription
consistent with the pure spinor constraint
(or more specifically,
with the gauge invariance under $\delta\omega_{\alpha}=(\gamma_{m}\lambda)_{\alpha}\Omega^{m}$).
In order to demonstrate how the prescription
regularizes the region $(\overline{\lambda}\lambda)\sim 0$,
we here ignore the subtleties coming from the pure spinor constraint.

\subsection{Terms requiring regularization of $(\overline{\lambda}\lambda)\sim0$}

To estimate the order of divergence
as $(\overline{\lambda}\lambda)$ approaches $0$,
it is more convenient to use $\omega_{\alpha}$
and $\overline{\omega}^{\alpha}$ instead
of their gauge invariant counterparts,
$(N_{mn},J_{\lambda})$ and $(\overline{N}_{mn},\overline{J}_{\lambda})$.
On a genus $g$ surface,
the zero mode integration measures
for the pure spinor variables behave as~\cite{Berkovits:2005bt}
\begin{align}
[D\lambda] & = \lambda^{-3}\mathd^{11}\lambda\,,\quad
[DN] = \lambda^{-8g}\mathd^g J\mathd^{10g}N
 \to [D\omega] = \lambda^{3g}\mathd^{11g} \omega\,,
\\
[D\overline{\lambda}] &= \overline{\lambda}^{-3}\mathd^{11}\overline{\lambda}\,,\quad
[D\overline{\omega}] = \overline{\lambda}^{3g}\mathd^{11g}\overline{\omega}\,,\quad
[D r]= \overline{\lambda}^{3}\mathd^{11}r \,,\quad
[D s] = \overline{\lambda}^{-3g}\mathd^{11g}s \,.
\end{align}
The total measure thus goes as
\begin{align}
[D\lambda D\overline{\lambda}D\omega D\overline{\omega} Dr Ds]
&= \lambda^{3g-3}\mathd^{22}\lambda\mathd^{22g}\omega \mathd^{11}r\mathd^{11g}s
\end{align}
where we denoted $\mathd^{22}\lambda = \mathd^{11}\lambda\mathd^{11}\overline{\lambda}$
etc.

So when the poles of the various operators in the correlator
add up to $\overline{\lambda}^{-11}\lambda^{-3g-8}$ or higher,
the zero-mode path integrals over $\lambda^{\alpha}$ and $\overline{\lambda}_{\alpha}$
become ill-defined because
\begin{align}
\label{eqn:lamlambarint}
\int\mathd^{22}\lambda\,{1 \over (\overline{\lambda}\lambda)^{L}}
\end{align}
diverges at $(\overline{\lambda}\lambda)\to0$ for $L\ge11$.
Fortunately, it turns out that
each factor of $(\overline{\lambda}\lambda)^{-1}$ comes with
a factor of $r_{\alpha}$, so for $L>11$
one always gets a zero from over-saturated $r_{\alpha}$ zero modes
as well. (The case of $L=11$ will be discussed at the end of section 
~\ref{subsec:regtwo}.)
To see this,
note that on a genus $g$ surface, one needs $3g-3$ $b$ ghost insertion
and each term in the $b$ ghost goes as
${\overline{\lambda}r^{k-1}/(\overline{\lambda}\lambda)^{k}}$.
Hence, after combining with the $\lambda^{3g-3}$ pole from the measure,
one indeed gets integration of the form
\begin{align}
\int\mathd^{22}\lambda\mathd^{11}r\,\sum_{k} {r^{k}\over(\overline{\lambda}\lambda)^{k}} \,.
\end{align}

Therefore, the problem is again to define the integral of
\begin{align}
\label{eqn:lamlambarint2}
\int\mathd^{22}\lambda\mathd^{11}r\,
{ r^{L}\over (\overline{\lambda}\lambda)^{L}}
 \sim 0\cdot \infty \,,\quad(L\ge11).
\end{align}
Using the idea described in~\cite{Berkovits:2006vi},
we now explain that an appropriate regularization
can be done by an insertion of an operator of the form
$\calN'(y)=\exp\{Q,\chi(y)\}=1+Q\Omega(y)$.

\subsection{Regularization of $(\overline{\lambda}\lambda)\sim0$}
\label{subsec:regtwo}

In order to concentrate on the main idea,
we first explain how to regularize
the divergent integral~(\ref{eqn:lamlambarint})
over $\lambda^{\alpha}$ and $\overline{\lambda}_{\alpha}$,
ignoring the zero that comes from $r_{\alpha}$ integration.
The method can be extended to respect the BRST invariance,
and then it naturally defines the $0\cdot\infty$ of~(\ref{eqn:lamlambarint2}).

The basic idea of~\cite{Berkovits:2006vi} is to
prevent the $(\overline{\lambda}\lambda)$ poles in operators
at different worldsheet positions
from diverging simultaneously.
This can be achieved by shifting the target-space location
of each pole by different constants $f_{i}$'s:
\begin{align}
\prod_{i=1}^{n}{1\over |\lambda(w_{i})|^{2l_{i}}}
\quad\to\quad
\prod_{i=1}^{n}{1\over |\lambda(w_{i})+f_{i}|^{2l_i}}\,.
\end{align}
Then, after integrating out the non-zero modes,
one is left with the integrand of the form
\begin{align}
\prod_{i=1}^{n}{1\over |\lambda+f_{i}|^{2l_i}}\,,\quad\sum l_{i} = L
\end{align}
instead of just ${(\overline{\lambda}\lambda)^{-L}}$.
The integration over $(\lambda,\overline{\lambda})$
is now well defined as long as $l_{i}<11$ in
each factor ${|\lambda+f_{i}|^{-2l_i}}$.
Eventually, one can average over the constants $f_{i}$'s
to get a finite result.

To achieve the shifting in the pure spinor formalism
in a BRST invariant manner,
we introduce a constant pure spinor $f^{\alpha}$
and its fermionic partner $g^{\alpha}$
(targetspace differential $\sim \mathd f^{\alpha}$),
as well as their complex conjugates
$\overline{f}_{\alpha}$ and $\overline{g}_{\alpha}$.
They are constrained as
\begin{align}
f^{\alpha}\gamma^{m}_{\alpha\beta}f^{\beta}
= f^{\alpha}\gamma^{m}_{\alpha\beta}g^{\beta}
= \overline{f}_{\alpha}\gamma_{m}^{\alpha\beta}\overline{f}_{\beta}
= \overline{f}_{\alpha}\gamma_{m}^{\alpha\beta}\overline{g}_{\beta}=0\,.
\end{align}
Then, we extend the BRST operator to
\begin{align}
Q \to Q'=Q+f^{\alpha}{\del\over \del g^{\alpha}}
  + \overline{g}_{\alpha}{\del \over \del \overline{f}_{\alpha}}
\end{align}
and introduce an additional regularization factor of the form
\begin{align}
\calN'(y) &= \exp\{Q,\chi(y)\}
\end{align}
at an arbitrary point $y$ on the worldsheet.
To achieve the shift, we include $g^{\alpha}\omega_{\alpha}(y)+\overline{f}_{\alpha}s^{\alpha}(y)$
in $\chi(y)$, and also put zero modes $\sum_{I}\omega_{\alpha,I}s^{\alpha}_{I}$
to impose an exponential cut-off for
the integration over
$\omega_{\alpha}$ and $\overline{\omega}^{\alpha}$.\footnote{%
In~\cite{Berkovits:2006vi}, 
the zero mode cut-off
$\exp\{Q,-\overline{f}g\}=\exp(-\overline{f}f-\overline{g}g)$
was included in $\calN'$ instead of the cut-off for
$\omega_{\alpha}$ and $\overline{\omega}^{\alpha}$,
$\exp\sum(-\overline{\omega}\omega-sd)$.
Since we have a factor of $\exp(f\omega+\overline{f}\overline{\omega})$
in $\calN'$ as well,
both should have the same effect upon integration, 
but we found it simpler in practice to use our choice.}
So we have
\begin{align}
\begin{split}
\calN'(y) &=
\exp[-\sum_{I=1}^{g}
  (\overline{\omega}^{\alpha}_{I}\omega_{\alpha,I}+s^{\alpha}_{I}d_{\alpha,I})]\\
&\qquad\times \exp(f^{\alpha}\omega_{\alpha}(y)+g^{\alpha}d_{\alpha}(y)
 + \overline{f}_{\alpha}\overline{\omega}^{\alpha}(y)+\overline{g}_{\alpha}s^{\alpha}(y)) \,.
\end{split}
\end{align}
Unlike the zero mode regulator $\calN_{0}$ for $(\overline{\lambda}\lambda)\to\infty$,
the regulator $\calN'(y)$ includes
non-zero modes in an essential way.
In particular, non-zero modes of $s^{\alpha}$ are important
for removing extra $r_{\alpha}$'s in the integrand.

The easiest way
to understand that $\calN'(y)$ indeed brings about
the desired shift of poles
is to use the path integral formalism.
On a genus $g$ surface, each component of $\omega_{\alpha}$ and $\lambda^{\alpha}$ can be
expanded by a complete set of eigenfunctions
of the worldsheet Laplacian as:
\begin{align}
\omega_{\alpha}(z) &= \sum_{I=1}^{g}\omega_{\alpha,I}\Omega_{I}(z)
  + \sum_{I'} \omega_{\alpha,I'}\Omega_{I'}(z,\zbar)
\\
\lambda^{\alpha}(z) &= \lambda^{\alpha}_{0}\Lambda_{0}
 + \sum_{I'}\lambda^{\alpha}_{I'}\Lambda_{I'}(z,\zbar)\,.
\end{align}
Here, $\Omega_{\alpha,I}(z)$ ($I=1\sim g$)
are $g$ zero modes of $\omega_{\alpha}$,
$\Lambda_{0}$ is the zero mode of $\lambda^{\alpha}$,
$\Omega_{I'}(z)$ and $\Lambda_{I'}(z)$ are the non-zero modes
of $\omega_{\alpha}$ and $\lambda^{\alpha}$.
Then, the Green function
\begin{align}
G(y,z)
 &= \sum_{I'}\Omega_{I'}(y,\ybar)\Lambda_{I'}(z,\zbar)
\end{align}
satisfies
\begin{align}
\del_{\ybar}G(y,z)
&= \sum_{I'}\Lambda_{I'}^{\ast}(y,\ybar)\Lambda_{I'}(z,\zbar)
 = \delta^{2}(z-y)-|\Lambda_{0}|^{2}
\\
\del_{\zbar}G(y,z)
&= \sum_{I'}\Omega_{I'}(y,\ybar)\Omega^{\ast}_{I'}(z,\zbar)
 = \delta^{2}(z-y)-\sum_{I}\Omega_{I}(y)\Omega^{\ast}_{I}(z) \,.
\end{align}

Now, using the Green function $G(y,z)$,
$\calN'(y)$ can be rewritten as
\begin{align}
\calN'(y)
&= \exp[-\sum_{I=1}^{g}(\overline{\omega}^{\alpha}_{I}
  \omega_{\alpha,I}+s^{\alpha}_{I}d_{\alpha,I})]\nonumber\\
&\qquad\times
\exp
\int\mathd^{2}z\delta^{2}(y-z)
 (f^{\alpha}\omega_{\alpha}+g^{\alpha}d_{\alpha}+\overline{f}_{\alpha}\overline{\omega}^{\alpha}
 +\overline{g}_{\alpha}s^{\alpha})(z) \\
&= \calN_{0}'
\times\exp{\int\mathd^{2}z \del_{\zbar}G(y,z)
 (f^{\alpha}\omega_{\alpha}+g^{\alpha}d_{\alpha}+\overline{f}_{\alpha}\overline{\omega}^{\alpha}
 +\overline{g}_{\alpha}s^{\alpha})(z)}\,,
\end{align}
where
\begin{align}
\calN_{0}'
&= \exp\sum_{I=1}^{g}
  [-(\overline{\omega}_{\alpha,I}\omega^{\alpha}_{I}+s^{\alpha}_{I}d_{\alpha,I})
 + (f^{\alpha}\omega_{\alpha,I}+g^{\alpha}d_{\alpha,I}+\overline{f}_{\alpha}\overline{\omega}_{I}^{\alpha}
  +\overline{g}_{\alpha}s^{\alpha}_{I})]\,,
\end{align}
and $(\omega_{\alpha,I},\overline{\omega}^{\alpha}_{I},d_{\alpha,I},s^{\alpha}_{I})$
are the $g$ zero modes of
$(\omega_{\alpha},\overline{\omega}^{\alpha},d_{\alpha},s^{\alpha})$.
Therefore, 
except for the zero mode factor $\calN_{0}'$,
insertion of $\calN'(y)$
in the path integral can be absorbed into
the change of variables\footnote{
We thank Joost Hoogeveen for pointing out that $x^m$ 
must also be included in the change of variables.}
\begin{align}
\label{eqn:shift}
\begin{split}
\lambda^{\prime \alpha}(z) &= \lambda^{\alpha}(z) + f^{\alpha}G(y,z)\,,\quad
\overline{\lambda}'_{\alpha}(z)
= \overline{\lambda}_{\alpha}(z) + \overline{f}_{\alpha}G(y,z)\,, \\
\theta^{\prime \alpha}(z) &= \theta^{\alpha}(z) + g^{\alpha}G(y,z) \,,\quad
r'_{\alpha}(z) = r_{\alpha}(z) + \overline{g}_{\alpha}G(y,z)\,, \\
x^{\prime m}(z) &= x^{m}(z) -{1\over 2}  g^{\alpha}(\gamma^m \theta (y))_\alpha
G(y,z) \,, \quad
d'_{\alpha}(z) = d_{\alpha}(z) + (g\gamma^m)_\alpha \Pi_m(y) G(y,z)\,, 
\end{split}
\end{align}
as in
\begin{align}
&\int\!D\phi \;\calN'(y)\exp(-S)
= \int\!D\phi\;\calN'_{0}\exp(-S')
\\
 S &= \int\mathd^{2}z
  ({1\over 2}
\del_z x^m \del_{\zbar} x_m + p\del_{\zbar}\theta
  - \omega\del_{\zbar}\lambda- \overline{\omega}\del_{\zbar}\overline{\lambda}
    +s\del_{\zbar}r)\,.
\end{align}
In other words,
the path integral definition of the correlation functions
\begin{align}
\langle O_{1}(w_{1})O_{2}(w_{2})\cdots O_{n}(w_{n})\rangle
&= \int D\phi \exp(-S)\calN'(y)O_{1}(w_{1})O_{2}(w_{2})\cdots O_{n}(w_{n})
\end{align}
is equivalent to
\begin{align}
\langle O_{1}(w_{1})O_{2}(w_{2})\cdots O_{n}(w_{n})\rangle'
&= \int D\phi' \exp(-S')\calN'_{0}(y)O'_{1}(w_{1})O'_{2}(w_{2})\cdots O'_{n}(w_{n}) \,,
\end{align}
where in $O'_{i}(w_{i})$,
variables $(\lambda^{\alpha},\overline{\lambda}_{\alpha},\theta^{\alpha},r_{\alpha})$
are shifted as in~(\ref{eqn:shift}).
For example, an operator of the form
\begin{align}
{r^{3} F(x,\theta) \over (\overline{\lambda}\lambda)^{3}}(w_{i})\,,\quad
\text{($F$: some superfield)}
\end{align}
gets modified to
\begin{align}
{(r'+{\overline{g}}_i)^{3} F(x'- {1\over 2}
 g_i\gamma\theta,\theta'+g_i) \over |\lambda'+f_{i}|^{6}}(w_{i})
\end{align}
where we abbreviated as
\begin{align}
f^{\alpha}_{i} &= f^{\alpha}G(y,w_{i})\,.
\end{align}
Below, we omit primes from $(\lambda^{\alpha},\overline{\lambda},\theta^{\alpha},r_{\alpha})$
for simplicity.

So, the computation of a $g$-loop amplitude
typically reduces to a sum of zero mode integrations of the form
\begin{align}
\label{eqn:zeromodeNprime}
\int\mathd^{16g}d\mathd^{16}\theta\mathd^{22g}\omega
\int \mathd^{22}f\mathd^{22}g\int\mathd^{22}\lambda\mathd^{11}r
\,\calN_{0}\calN'_{0}
  \prod_{i}{(r+{\overline{g}}_i)^{l_i}F_{i}(x-{1\over 2} g_i\gamma\theta,
\theta+g_i)\over |\lambda+f_{i}|^{2l_{i}}}\,.
\end{align}
When the total order of divergence $\sum l_{i}$ is smaller than $11$,
$\calN'_{0}$ regularization does nothing
and the integral goes back to that of section~\ref{sec:review}.
The case when $\sum l_{i}=11$ is special and will be discussed
at the end of this subsection.
When $11<\sum l_{i}<22$,
only the $\calN'_{0}$ regularization is necessary,
and we can set $\calN_{0}=1$ using BRST invariance.
Moreover, provided each $l_{i}$ is smaller than $11$,
one can show that the integral~(\ref{eqn:zeromodeNprime}) is finite
(possibly zero due to a lack of some fermionic zero modes).

To prove this, first introduce parameters 
$(c,\varepsilon,\overline{\varepsilon})$ in ${\cal N}'$ as
\begin{align}
\begin{split}
\calN'(y) &=
\exp[-c\sum_{I}(\overline{\omega}^{\alpha}_{I}\omega^{\alpha}_{I}+s^{\alpha}_{I}d_{\alpha,I})]\\
&\qquad\times \exp[\varepsilon(f^{\alpha}\omega_{\alpha}(y)+g^{\alpha}d_{\alpha}(y) +\overline{\varepsilon}(
\overline{f}_{\alpha}\overline{\omega}^{\alpha}(y)+
\overline{g}_{\alpha}s^{\alpha}(y))] \,.
\end{split}
\end{align}
BRST invariance guarantees that the scattering amplitude will
be independent of 
$(c,\varepsilon,\overline{\varepsilon})$. Performing
the integration over
the zero modes $\omega_{\alpha,I}$ and $\overline{\omega}^{\alpha}_{I}$
($I=1\sim g$),
this yields an exponential cut-off
\begin{align}
\exp[-{\varepsilon\overline{\varepsilon}\over c}
  (\overline{f}_{\alpha}f^{\alpha}+\overline{g}_{\alpha}g^{\alpha})]
\end{align}
for $f^{\alpha}$ and $\overline{f}_{\alpha}$.
Then, it can be shown that
the integration over $(\lambda^{\alpha},\overline{\lambda}_{\alpha},f^{\alpha},\overline{f}_{\alpha})$
is finite.
When $\sum l_{i}>11$, it is clear that
the only region of integration that can cause divergence
is where all $(\lambda^{\alpha},\overline{\lambda}_{\alpha},f^{\alpha},\overline{f}_{\alpha})$
become small of order $\epsilon$.
But thanks to the smearing by $f$'s,
the integral is now finite in this region for $11<L=\sum l_{i}<22$:
\begin{align}
\label{eq:divone}
\int\mathd^{22}\lambda{1\over|\lambda|^{2L}}
\sim \epsilon^{22-2L}
\quad\to\quad
\int\mathd^{22}f\mathd^{22}\lambda{1\over|\lambda+f|^{2L}}
\sim \epsilon^{44-2L}\,.
\end{align}

Although the $\calN'$ regularization described so far
only works when the order of $(\overline{\lambda}\lambda)$ is below $22$,
the estimate of the integral above also shows what to do
if the order of $(\overline{\lambda}\lambda)$ poles sum up beyond $22$.
For example, when the total order of divergence $L$
satisfies $22<L<33$,
one only has to introduce another copy of
smearing variables
$(f^{\prime \alpha},g^{\prime \alpha},\overline{f}'_{\alpha},\overline{g}'_{\alpha})$,
and extend the regulator $\calN'(y)$
so that $\lambda^{\alpha}$ gets shifted by both $f^{\alpha}$ and $f^{\prime \alpha}$.
Then, one eventually gets the integral of the form
\begin{align}
\label{eq:divtwo}
\int\mathd^{22}f\mathd^{22}f'\mathd^{22}\lambda{1\over|\lambda+f+f'|^{2L}}
\sim \epsilon^{66-2L}\,,
\end{align}
which is finite when $L<33$.

Finally, we shall discuss the case when $\sum l_i= 11n$ for any positive
integer $n$ and will argue that these terms do not contribute to
BRST-invariant amplitudes. In this case,
an analysis similar to~(\ref{eq:divone}) and~(\ref{eq:divtwo})
implies that integration over the bosonic
variables $(\lambda^\alpha,f^\alpha)$ gives a logarithmic dependence on
$\varepsilon$. However, integration over the fermionic variables 
$(\theta^\alpha, g^\alpha)$ can only give polynomial dependence on 
$\varepsilon$. Since BRST-invariant amplitudes must be independent
of $\varepsilon$, this suggests that terms with $\sum l_i=11n$
cannot contribute to BRST-invariant scattering amplitudes.

To see more explicitly why this happens, consider the term with
$\sum l_i =11$ 
\begin{align}
V_{11}\equiv {{(\lambda^3 r^{11})}\over{({\overline\lambda}\lambda)^{11}}} 
(\theta)^{16}
\end{align}
where $(\lambda^3 r^{11})$ denotes the unique Lorentz-invariant contraction
of three pure spinor $\lambda$'s and 11 fermionic $r$'s.
Since the functional integral 
\begin{align}
\int\mathd^{22}\lambda\mathd^{11}r\,\mathd^{16}\theta\,{\cal N}\, V_{11}
\end{align}
is nonzero, such a term would contribute to the scattering amplitude if
$V_{11}$ were present in the BRST invariant
integrand $f(\lambda,\overline\lambda, r,\theta)$
of~(\ref{eqn:leftone}).

However, $V_{11}$ cannot appear as part of a BRST-invariant
expression for the following reason. First,
note that $V_{11}$
is in the cohomology of $Q_1 = \oint r_\alpha {\overline \omega}^\alpha$
and that the only non-trivial cohomology
of $Q=Q_{0}+Q_{1}$ at weight $0$ and ghost number $3$ is
$(\lambda^3\theta^{5})\equiv(\lambda\gamma^{m}\theta)(\lambda\gamma^n\theta)(\lambda\gamma^{p}\theta)(\theta\gamma_{mnp}\theta)$.
So the only possible way for $V_{11}$ to be a part of
a BRST invariant operator $\sum_{k=0}^{11} V_{k}$ is if 
\begin{align}
\sum_{k=0}^{11} V_{k}=a(\lambda^{3}\theta^{5})+Q(\sum_{k=0}^{11}\Lambda_{k})
\end{align}
for some set of $\Lambda_{k}$'s,
where $a$ is a constant and the subscript $k$ denotes the term proportional
to $r^k$. Since $Q=Q_0+Q_1$ where $Q_0 = \lambda^\alpha 
{\del\over{\del\theta^\alpha}}$ and $Q_1 = r_\alpha
{\del\over{\del {\overline\lambda}_\alpha}}$,
this would
imply that $V_{11}=Q_0 \Lambda_{11} + Q_1 \Lambda_{10}$.
Since $V_{11}$ is proportional to $(\t)^{16}$ and $Q_0$ lowers
the number of $\t$'s, $\Lambda_{11}$ must vanish which implies
that $V_{11}=Q_1\Lambda_{10}$. 
But this cannot happen because 
$V_{11}$ is in the cohomology of $Q_{1}$.
Therefore, the eleventh pole simply does not contribute
to the BRST invariant amplitude,
and the $\calN'$ regularization is unnecessary.

Similarly, the only dangerous term for $\sum_{i}l_{i}=22$
with a single set of smearing variables is
\begin{align}
V_{22} \equiv {(\lambda^{3}r^{11})(\overline{g}g)^{11}(\theta)^{16}
 \over \prod_{i}|\lambda+f_{i}|^{2l_{i}}}
 \in {(\lambda)^{3}(r+\overline{g})^{22}(g)^{11}(\theta)^{16}
 \over \prod_{i}|\lambda+f_{i}|^{2l_{i}}} \,.
\end{align}
Again, one can argue that this is in the cohomology of
$Q_{1}=r_{\alpha}
 {\del\over\del\overline{\lambda}_{\alpha}}
 +\overline{g}_{\alpha}{\del\over\del\overline{f}_{\alpha}}$
but cannot be part of a BRST invariant operator
where now the BRST operator is extended to $Q=Q_{0}+Q_{1}$
with $Q_{0}=\lambda^{\alpha}
{\del\over\del \theta^{\alpha}}
+f^{\alpha}{\del\over\del g^{\alpha}}$.
So the twenty-second pole cannot contribute to the
BRST invariant amplitude as well.

\bigskip
It is important to stress that the $\calN'(y)$ regulator is
necessary for resolving $0/0$ ambiguities and does not affect
the computation of terms with $\sum l_i< 11$ where such ambiguities
are not present. Similarly, if one has two copies of 
$\calN'(y_1) \calN'(y_2)$ inserted at different points
of the worldsheet, the extra regulator does not affect the computation of
terms with $\sum l_i< 22$. So for an amplitude computation whose maximum
order of divergence is $L=\sum l_i$, it is sufficient to insert $L/11$
regulators at different points on the worldsheet. Of course, one can
always insert more than $L/11$ regulators, but the additional regulators
will not affect the computation.

It is also important to stress that if one does not include a sufficient
number of regulators, the amplitude is ambiguous since one has
a divergence (coming from the bosonic functional integral) multiplied
by a zero (coming from the fermionic functional integral). In a recent
paper \cite{grassi:2009gv}, Grassi and Vanhove claimed that the 
amplitude is vanishing if one does not include a sufficient number of
regulators, which would lead to a violation of unitarity. Their claim
was based on doing the fermionic functional integral {\it before}
doing the bosonic functional integral, in which case they obtained the
zero but ignored the divergence. However, any proper regularization procedure
should be independent of the order of integration. So it is necessary to
first regularize the divergence in the
bosonic functional integral before attempting to do
the fermionic functional integral. 

Grassi and Vanhove also proposed a new regulator in \cite{grassi:2009gv}
of the form
\begin{align}
\widehat{\calN} = \exp[-{1\over{\lambda\bar\lambda}} -
r_\alpha ({{\delta^\alpha_\beta}\over{(\lambda\bar\lambda)^2}}
-2 {{\lambda^\alpha\bar\lambda_\beta}\over{(\lambda\bar\lambda)^3}})
\theta^\beta]
\end{align}
which removes the divergence when $\lambda\bar\lambda \to 0$.
However, their regulator does not remove the divergence when 
$\lambda\bar\lambda \to \infty$.
Although they claim that the divergence when
$\lambda\bar\lambda \to \infty$ can be ignored because of
the $(\lambda\bar\lambda)^{-1}$ factors coming from the integration
over the $r$ zero modes, this claim is based on the assumption that
one can first do the fermionic functional integral before doing the bosonic
functional integral. However, as stressed above, any proper regularization
procedure must be independent of the order of integration.
Since performing
the bosonic functional integral {\it before} performing
the integration over the $r$ zero modes leads to an ambiguous answer
using the regulator of
\cite{grassi:2009gv}, their regularization procedure is incomplete.  
Regularization of the
bosonic functional integral when $\lambda\bar\lambda\to\infty$
is expected to require additional insertions
involving $d$ zero modes. Note that these extra
$d$ zero modes must arise from the regularization of divergences, and
cannot be included by simply modifying the gauge-fixing condition as suggested
in \cite{grassi:2009gv}.

\bigskip
This concludes our explanation of the regularization method
of~\cite{Berkovits:2006vi}
and we now turn to the construction
of vertex operators in the Siegel gauge.

\section{Vertex operators in the Siegel gauge}
\label{sec:siegel}

in the Siegel gauge.
In the Siegel gauge, vertices are annihilated by $b_0$
so the equation of motion implied by $Q=0$
is simply $L_0=0$ (and $\Box=0$ for the massless vertex).
Therefore, the gauge is an extension of the Lorentz gauge
where the photon wave function satisfies $P^m a_m (x)=0$.

\subsection{Unintegrated vertex operators in the Siegel gauge}

In the review of the pure spinor formalism above,
we defined the physical vertex operators
to be the cohomology elements of the BRST operator
\begin{align}
 Q &= Q_0 + Q_1\,,
\end{align}
where $Q_{0} = \oint\lambda^{\alpha}d_{\alpha}$,
and $Q_1 = \oint r_{\alpha}\overline{\omega}^{\alpha}$.
Conventionally, however, the vertex operators
have been assumed to be annihilated by $Q_0$ and $Q_1$
separately.\footnote{
In the \v{C}ech type formulation of curved $\beta\gamma$ systems,
this corresponds to the operators
defined globally on the pure spinor space.}
This is a particular choice of a gauge for the cohomology representatives
of $Q=Q_0+Q_1$;
we shall refer to this gauge as the ``minimal gauge''.

In the minimal gauge, vertices are independent of
non-minimal variables,
$(\overline{\omega}^{\alpha},\overline{\lambda}_{\alpha},s^{\alpha},r_{\alpha})$,
and have no poles in $\overline{\lambda}\lambda$.
For example,
the unintegrated vertex operators for
the massless and the first massive modes are
given by the most general
$\lambda$-charge $1$ operators of this type,
\begin{align}
\label{eq:globalVmassless}
V^{\text{min}}_{\text{massless}}
&= \lambda^{\alpha}A_{\alpha}(x,\theta) \,, \\
V^{\text{min}}_{\text{1st massive}}
&= \del\lambda^{\alpha}B_{\alpha}(x,\theta)
 + \lambda^{\alpha}\del\theta^{\beta}B_{\alpha\beta}(x,\theta)
 + \lambda^{\alpha}d_{\beta}B_{\alpha}{}^{\beta}(x,\theta) \nonumber\\
&\qquad + \lambda^{\alpha}\Pi^{m}B_{\alpha m}(x,\theta)
 + \lambda^{\alpha}J B_{\alpha}(x,\theta)
 + \lambda^{\alpha}N^{mn}B_{\alpha mn}(x,\theta) \,,
\end{align}
and it had been explicitly checked
that the cohomology condition with respect to
$Q_{0}=\oint\lambda^{\alpha}d_{\alpha}$
yields the correct on-shell constraints
and gauge invariance conditions on the superfields
$A_{\alpha}(x,\theta)$, $B_{\alpha}(x,\theta)$, $\cdots$,
$B_{\alpha mn}(x,\theta)$~\cite{Berkovits:2000fe,Berkovits:2002qx}.

Although the minimal gauge is convenient for many calculations,
it is more natural and sometimes necessary
to allow more general dependencies on the
pure spinor sector.
A natural class of (massless) vertex operators
with ghost number $1$ is given by
\begin{align}
\label{eq:generalVmassless}
\begin{split}
V &= V_1 + \cdots + V_p \\
&= {\overline{\lambda}_{\alpha_0}\lambda^{\beta}\lambda^{\gamma}
  C^{\alpha_0}{}_{\beta\gamma}(x,\theta)\over (\lambda\overline{\lambda})}
 + \cdots +
 {\overline{\lambda}_{\alpha_0}r_{\alpha_1}\cdots r_{\alpha_p}\lambda^{\beta}\lambda^{\gamma}
  C^{\alpha_0\alpha_1\cdots \alpha_p}{}_{\beta\gamma}(x,\theta)
 \over (\lambda\overline{\lambda})^{p+1}}\,,
\end{split}
\end{align}
where the ghost number is measured by $J_{g}=\omega\lambda-\overline\omega
\overline\lambda$.
Note that as in the $b$ ghost,
$V$ has been defined such that 
$(r_{\alpha} {\del\over{\del r_{\alpha}}} +  
{\overline\lambda}_{\alpha} {\del\over{\del {\overline\lambda}_{\alpha}}})V=0.$ 
Since the composite $b$-ghost depends non-trivially
on $r/(\lambda\overline{\lambda})$,
it is clear that the Siegel gauge condition
$b_0=0$ can be achieved only if one allows
the vertex operators to depend on the non-minimal fields
as in~(\ref{eq:generalVmassless}).

Note that the choice of~(\ref{eq:generalVmassless})
allows one to choose different gauges for the superfields $C$'s
on different coordinate patches of the pure spinor space.
The simple form of the vertex in~(\ref{eq:globalVmassless})
should then be understood as a special choice of the gauge for $C$'s
such that the vertex is globally defined on the pure spinor space
(or in other words, independent of the non-minimal variables).

We shall now show that the Siegel gauge can
be achieved within the general form of the vertices~(\ref{eq:generalVmassless})
by explicitly constructing them.
To be concrete, we explain our construction
using the massless vertex as an example,
but the construction works for massive fields as well.\footnote{
Antifield vertices for the massive modes have not been
computed explicitly in the pure spinor literatures,
but strong evidence for the fact
that the space of pure spinor vertices enjoys
field-antifield symmetry was presented in~\cite{Aisaka:2008vw,Aisaka:2008vx}.}
We start from the ghost number $2$ cohomology of $Q$
\begin{align}
  V^{\ast} &= \lambda^{\alpha}\lambda^{\beta}A_{\alpha\beta}(x,\theta)\,,
\end{align}
with $A_{\alpha\beta}(x,\theta)$ a {\em bosonic} superfield.
Our Siegel gauge vertex operator is then defined as
\begin{align}
 V(z) &= b_0 V^{\ast}(z) \equiv \oint\mathd y (y-z)b(y)V^{\ast}(z)
\end{align}
which is obviously annihilated by $b_0$.
More explicitly, $V$ reads
\begin{align}
 \label{eqn:MasslessSiegel}
V  & = V_{0} + V_{1} + V_{2} + V_{3}\,,
\end{align}
where
\begin{align}
V_0 
&= (b_{0})_{0}V^{\ast}
 = -{\overline{\lambda}_\alpha\lambda^\beta\lambda^\gamma (\fslash{P}D)^\alpha A_{\beta\gamma} \over 2(\overline{\lambda}\lambda)}\,,\\
V_1 
&= (b_{1})_{0}V^{\ast}
= {(\overline{\lambda}\gamma^{mnp}r)[\lambda^\beta\lambda^\gamma (D\gamma_{mnp}D)-24(\gamma_{mn}\lambda)^\beta\lambda^\gamma P_{p}]A_{\beta\gamma}
 \over 192(\overline{\lambda}\lambda)^2}\,,
 \\
V_{2}
&= (b_{2})_{0}V^{\ast}
=-{\overline{\lambda}_\alpha(r\gamma^{mnp}r)(\gamma_{mn}\lambda)^\beta\lambda^\gamma(\gamma_{p}D)^\alpha A_{\beta\gamma}
 \over 16(\overline{\lambda}\lambda)^3}\,,
\\
V_{3}
&= (b_{3})_{0}V^{\ast}
= {(\overline{\lambda}\gamma^{mnp}r)(r\gamma_{p}{}^{qr}r)(\gamma_{mn}\lambda)^\beta(\gamma_{qr}\lambda)^\gamma A_{\beta\gamma}
 \over 256(\overline{\lambda}\lambda)^4}\,.
\end{align}

Let us now study the gauge condition implied by
the construction above.
The cohomology condition on $V^{\ast}$ implies that $A_{\alpha\beta}(x,\theta)$
is subject to the constraint and gauge invariance of
\begin{align}
 D_{\lpar \alpha}A_{\beta\gamma \rpar} &= 0\,,\quad
 \delta A_{\alpha\beta} = D_{\lpar\alpha}\Omega_{\beta\rpar}\,,
\end{align}
where $\Omega_{\beta}(x,\theta)$ is an arbitrary superfield,
and the notation $\lpar\alpha_{1}\cdots \alpha_{n}\rpar$ signifies
symmetric $\gamma$-traceless combination of spinorial indices.
From the antifield calculus in ten dimensional
super-Maxwell theory,
it is well known~\cite{Berkovits:2001rb,Cederwall:2001dx} that
$A_{\alpha\beta}(x,\theta)$ contains a vector at $\theta^{4}$,
whose ``equation of motion'' is the Lorentz gauge condition
\begin{align}
 P^{m}a_{m}(x) = 0 \,.
\end{align}
So our construction explains that the Siegel gauge
$b_{0}V=0$ is indeed an extension of the Lorentz gauge,
as is expected.

\bigskip
Several remarks are in order
before turning to the construction of integrated
vertex operators in the Siegel gauge.

First, we note that the construction above
in fact parallels that of the bosonic string.
In the bosonic string, the notion of the Siegel gauge
and that of field ($V=c\psi(x)$)
and antifield ($V^{\ast}=c\del c\psi^{\ast}(x)$)
are manifestly related because there
the field-antifield doubling comes from the ghost zero-mode
oscillators satisfying $\{b_{0},c_{0}\}=1$.
Also, it is clear that the field (or Siegel gauge) vertex operators
can be obtained from the antifield vertex operators
by acting with $b_{0}$.

In the pure spinor formalism, however, there is no $c$ ghost
so {\it a priori}
the field-antifield doubling and the Siegel gauge choice
are unrelated.
Moreover, the $b$ ghost is a complicated operator
that may have non-trivial cohomologies,
so one might even worry that the condition $b_0 V=0$
on the vertex of the form~(\ref{eq:generalVmassless})
does not have a solution.
However, our construction explains that the only structure
needed for solving $b_0V=0$ is the field-antifield symmetry
of the operator space.
The presence of the field-antifield symmetry is non-trivial
in the pure spinor formalism, but is strongly supported by
the study of the pure spinor partition functions
in~\cite{Aisaka:2008vw,Aisaka:2008vx}.

Finally, in our construction, we have not specified the gauge for
the superfield in $V^{\ast}$.
However, it is easy to see that any choice of gauge
leads to a vertex in the Siegel gauge.
The ``pre-gauge transformation''
$\delta V^{\ast} = Q(\lambda^{\alpha}\Omega_{\alpha})$
simply modifies the Siegel gauge vertex $V$ by
\begin{align}
  \delta V &= b_{0}(Q(\lambda^{\alpha}\Omega_{\alpha})) = L_{0}(\lambda^{\alpha}\Omega_{\alpha})
  - Q(b_{0}(\lambda^{\alpha}\Omega_{\alpha}))
\end{align}
The first term vanishes if $\Omega_{\alpha}$ has weight $0$,
and the second term is the remaining gauge transformation in Siegel gauge
analogous to the residual gauge transformation
of the Maxwell theory in the Lorentz gauge.

\subsection{Massless integrated vertex operator in the Siegel gauge}

We now turn to the construction of the integrated vertex operators.
We exclusively consider massless vertices.
Given a $Q$-closed unintegrated vertex operator $V$
in an arbitrary gauge, the
corresponding integrated vertex operator ${U}$
satisfying
\begin{align}
 \label{eq:unintegprop}
  Q{U}(w) = \del V(w)
\end{align}
can be obtained by defining
\begin{align}
  {U}(w) = b_{-1}V(w)\,.
\end{align}
Since $\{Q\,, b_{-1}\}=L_{-1}$,
it is clear that ${U}$ satisfies~(\ref{eq:unintegprop}).

For the Siegel gauge vertex operator $V=b_0 V^{\ast}$
of the previous subsection,
${U}$ is a conformal primary of weight $1$.
Indeed, since $b$ and $V^{\ast}$ has at most
a double pole, 
one easily finds that for $n>0$,
\begin{align}
  L_{n}{U} = b_{n-1}(b_{0}V^{\ast}) + b_{-1}(b_{n}V^{\ast})
  = -b_{0}(b_{n-1}V^{\ast}) +0 = 0\,.
\end{align}

Schematically, the integrated vertex operator $U=b_{-1}V$ is
of the form
\begin{align}
U &= (b_{-1})_{-1}V
 + \del\theta^{\alpha}f_{\alpha}
 + \Pi^{m}f_{m}
 + d_{\alpha}f^{\alpha}
 + {1\over2}N^{mn}f_{mn}\,,
\end{align}
where $(b_{-1})_{-1} V$ denotes the simple pole of
$(s^\alpha \del\overline\lambda^\alpha)$ with $V$, and the 
$f$'s are constructed
from $\lambda^{\alpha}$, $\overline{\lambda}_{\alpha}$, $r_{\alpha}$,
and spacetime derivatives of the superfield $A_{\alpha\beta}$,
e.g.\ $D^{n}A_{\alpha\beta}(x,\theta)$.
Since their $r$ dependence and the order of divergence
as $(\overline{\lambda}\lambda)\to 0$ become important for our application,
we record them here:
\begin{align}
f_{\alpha}
&= { D^{3}A_{\alpha\beta}\over(\overline{\lambda}\lambda)^{0}}
+ {r D^{2}A_{\alpha\beta}\over(\overline{\lambda}\lambda)^{1}}
+ \cdots
+{r^{3} A_{\alpha\beta}\over(\overline{\lambda}\lambda)^{3}}\,,
\\
f_{m} &= { D^{4}A_{\alpha\beta}\over(\overline{\lambda}\lambda)^{0}}
+ {r D^{3}A_{\alpha\beta}\over(\overline{\lambda}\lambda)^{1}}
+ \cdots
+ { r^{4} A_{\alpha\beta}\over(\overline{\lambda}\lambda)^{4}}\,,
\\
f^{\alpha}
&= { D^{5}A_{\alpha\beta}\over(\overline{\lambda}\lambda)^{0}}
+ {r D^{4}A_{\alpha\beta}\over(\overline{\lambda}\lambda)^{1}}
+ \cdots
+{r^{5}  A_{\alpha\beta}\over(\overline{\lambda}\lambda)^{6}}\,,
\\
f_{mn} &= { D^{6}A_{\alpha\beta}\over(\overline{\lambda}\lambda)^{0}}
+ {r  D^{5}A_{\alpha\beta}\over(\overline{\lambda}\lambda)^{1}}
+ \cdots
+{r^{6} A_{\alpha\beta}\over(\overline{\lambda}\lambda)^{6}}\,.
\end{align}

Although the vertex operator $U$ appears complicated,
it simplifies considerably after using the gauge invariance
$\delta A_{\alpha\beta} = D_{\lpar\alpha} A_{\beta\rpar}$ to gauge-fix
\begin{align}
(\overline{\lambda}\gamma^{m})^{\alpha}A_{\alpha\beta}=0.
\end{align}
To see that this gauge choice is accessible, choose a $U(1)\times SU(5)$
decomposition of $SO(10)$ such that the only non-vanishing component
of $\overline\lambda_\alpha$ carries $-{5\over 2}$ $U(1)$ charge.
If $A_{ab}$ (for $a=1$ to 5) denotes the component of 
$A_{\alpha\beta}$ with $+3$
$U(1)$ charge, the constraint $\lambda^\alpha \lambda^\beta
\lambda^\gamma D_\alpha A_{\beta\gamma} =0$ implies that
$D_{(a} A_{bc)}=0$ where $D_a$ is the component of $D_\a$ with $3\over 2$
$U(1)$ charge. Since $\{D_a , D_b\}=0$, 
$D_{(a} A_{bc)}=0$ implies that $A_{ab} = D_{(a}\Omega_{b)}$ for
some $\Omega_b$. So $\Omega_b$ can be used to gauge $A_{ab}=0$.
In the gauge $A_{ab}=0$, 
$\lambda^\alpha \lambda^\beta
\lambda^\gamma D_\alpha A_{\beta\gamma} =0$ implies that
$D_{(a} A_{b)}^{[cd]} =0$ where $A_b^{[cd]}$
denotes the component of $A_{\alpha\beta}$ with $+1$
$U(1)$ charge. So $A_b^{[cd]} = D_b \Omega^{[cd]}$ for some 
$\Omega^{[cd]}$, which means that $A_b^{[cd]}$ can also be gauged
to zero. In the gauge where $A_{ab}=A_b^{[cd]}=0$, it is easy
to verify that 
$(\overline{\lambda}\gamma^{m})^{\alpha}A_{\alpha\beta}=0$.

Since $(\overline\lambda\gamma^m r)=0$ implies
that the $U(1)$ charge of $r_\alpha$ is either $-{1\over 2}$ or
$-{5\over 2}$, one can use $U(1)$ invariance to verify in this gauge that
all terms beyond $r^{3}$ in $(f_{\alpha},f_{m},f^{\alpha})$ vanish,
and that all terms beyond $r^{4}$ in $f_{mn}$ vanish. Note that
$\lambda^\alpha \lambda^\beta
\lambda^\gamma D_\alpha A_{\beta\gamma} =0$ implies in this gauge that
the $U(1)$ charge of 
$D_\alpha A_{\beta\gamma}$ is less than or equal to $-{3\over 2}$.
It will turn out that when we compute
$4$-point $1$-loop amplitude using $4$ $U$'s,
the only contribution will come from
the $r^{3}$ term in $d_{\alpha}f^{\alpha}$
and the $r^{4}$ term in ${1\over2}N^{mn}f_{mn}$,
namely
\begin{align}
 d_{\alpha}{r^{3}D^{2}A_{\alpha\beta}(x,\theta)\over(\overline{\lambda}\lambda)^{3}}
+ N_{mn} {r^{4}D^{2}A_{\alpha\beta}(x,\theta)\over(\overline{\lambda}\lambda)^{4}}
\,.
\end{align}

This concludes our construction of the
massless integrated vertex operator
in the Siegel gauge, and we now argue
that it can be used to compute $n$-point $1$-loop
amplitudes using only integrated vertex operators.

\section{New $n$-point $1$-loop amplitude prescription}
\label{sec:application}

In this section, it will be shown that 
$n$-point $1$-loop amplitudes in the pure spinor formalism
can be computed using $n$ Siegel gauge
integrated vertex operators
of the previous section.

\subsection{Description of the problem}

In bosonic string theory,
the canonical prescription for computing $n$-point $1$-loop
amplitudes is to use $1$ unintegrated vertex operator
and $n-1$ integrated operators, with a single insertion
of the $b$ ghost:
\begin{align}
\label{eq:bosonic1loopA}
\calA_{n}
&= \int\mathd^2\tau
 \int\Bigl(\prod_{i=2}^{n}\mathd^2w_i\Bigr)
\bigl|\langle\int\!\mathd^2z (b\cdot \mu)(z)
  V(w_1)\prod_{i=2}^{n}U(w_i)
 \rangle\bigr|^2
\,.
\end{align}
However, it is well known that when the vertex operators are
in Siegel gauge, the amplitude
can also be computed using only the integrated vertex operators as in
\begin{align}
\label{eq:bosonic1loopB}
\calA_{n}
&= \int{\mathd^2\tau\over \Im \tau}
 \int\Bigl(\prod_{i=1}^{n}\mathd^2w_i\Bigr)
\bigl|\langle J_{g}(z)
  \prod_{i=1}^{n}U(w_i)
 \rangle\bigr|^2
\,.
\end{align}
Here $J_{g}=-bc$
is the ghost number current (put at an arbitrary
point $z$ on the worldsheet).
So a natural question is if a similar
prescription can also be used in the pure spinor formalism when
the vertex operators are in Siegel gauge.

To understand why Siegel gauge is necessary,
let us first explain why prescriptions of the type~(\ref{eq:bosonic1loopB})
with integrated vertex operators in the minimal gauge ($Q_{1}=0$)
give zero
for the massless $4$-point $1$-loop amplitude.
In the non-minimal formalism, the bosonic prescription~(\ref{eq:bosonic1loopB})
naively generalizes to
\begin{align}
\label{eq:nmpsBnaive}
\calA_{n}
&= \int{\mathd^2\tau\over \Im \tau} \int\mathd^2w_1 \cdots \mathd^2 w_n
\bigl|\langle \calN J_{g}(z)
  U(w_1)\cdots U(w_n)
 \rangle\bigr|^2
\,,
\end{align}
where $J_{g} = \omega\lambda -\overline \omega\overline\lambda$
is the ghost number current defined so that
the BRST charge $Q$ carries charge $+1$,
and
\begin{align}
\calN_{0} &= \exp[-(\overline{\lambda}\lambda+r\theta)
  -({1\over2}\overline{N}^{mn}N_{mn}+\overline{J}_{\overline{\lambda}}J_{\lambda}
  +{1\over2}S^{mn}(\lambda\gamma_{mn}d)+S(\lambda d)]
\end{align}
is the zero-mode regularization factor
that is needed to define
an indefinite factor ($\infty\cdot 0$)
coming from non-compact bosonic integrals
and unsaturated fermionic integrals.

Now, in order to have a non-vanishing result,
one must saturate the $16$ zero-modes of $d_{\alpha}$
on the torus.
However, $\calN_{0}$ can provide at most  $11$ $d_{\alpha}$ zero-modes,
and each unintegrated vertex operator can only provide
$1$ $d_{\alpha}$ zero-mode.
So, for the $4$-point amplitude
it is impossible to saturate the $d_{\alpha}$ zero-modes
in~(\ref{eq:nmpsBnaive}) and one gets a vanishing result.

To have a non-vanishing $4$-point $1$-loop amplitude
using $4$ integrated vertex operators,
an additional $d_{\alpha}$ zero-mode must be supplied from somewhere.
In~\cite{Berkovits:2006vi} it was suggested that the extra $d_{\alpha}$ zero-mode
could be provided from the additional regulator $\calN'(y)$
of section~\ref{sec:regularization},
that is needed when the total $\lambda\overline{\lambda}$ pole
in the integrand adds up greater than or equal to $11$.

Below, we shall show that in Siegel gauge, the
$1$-loop prescription of the form~(\ref{eq:bosonic1loopA})
with a $b$-ghost insertion can be converted to the
prescription of the form~(\ref{eq:bosonic1loopB})
that uses only the integrated vertex operators.
Moreover, we shall show that the additional regulator
of~\cite{Berkovits:2006vi}
does provide the missing $d_{\alpha}$ zero-mode so that
the $4$-point $1$-loop amplitude with this new prescription
is non-vanishing.

\subsection{The new $1$-loop prescription and its derivation}
\label{subsec:prescription}

In this subsection, we shall argue that the $n$-point $1$-loop
amplitude can be computed by the prescription of the form
\begin{align}
\label{eq:PS1loopB}
\calA_{n}
&= \int{\mathd^{2}\tau \over \Im\tau}
\Bigl(\prod_{i=1}^{n}\int\mathd^{2}w_{i} \Bigr)
  \bigl|\langle\calN_{0}\calN'_{0}
    \oint_{A}\!\mathd z J_{g}(z)
     \Bigl( \prod_{i=1}^{n} U'(w_{i})\Bigr)\rangle
  \bigr|^{2} \,,
\end{align}
where $\calN_{0}$ and $\calN'_{0}$
are the zero mode regularization factors reviewed above,
$J_{g}=\omega\lambda-\overline\omega\overline\lambda$ is the ghost number current,
and $U'$ is the ``smeared version'' of the
Siegel gauge integrated vertex operator.
The smearing was caused by the non-zero modes in $\calN'(y)$.
Below, we shall omit the prime (that denotes the smearing)
from various operators with its presence understood.

We start from the conventional prescription of the form
\begin{align}
\label{eq:PS1loopA}
\calA_{n}
&= \int\mathd^{2}\tau
\Bigl(\prod_{i=2}^{n}\int\mathd^{2}w_{i} \Bigr)
  \bigl|\langle\calN_{0}\calN'_{0}
    \oint_{A}\mathd z ( b\cdot \Delta v)(z)V(w_1)
     \Bigl( \prod_{i=2}^{n} U(w_{i})\Bigr)\rangle
  \bigr|^{2} \,.
\end{align}
Here, $V$ is the Siegel gauge unintegrated vertex $V=b_0V^{\ast}$
and ${U}=b_{-1}V$;
a non-trivial cycle $A$ and the discontinuity $\Delta v^{z}$ across $A$
of a quasi-conformal vector field $v^{z}$
are defined in a pair;
we take $A$ as a horizontal cycle of length $1$
on the real axis, and
\begin{align}
v^z={1\over(2\mathi\Im\tau)}(z-\zbar)
\end{align}
has a unit discontinuity across $A$.
$v^z$ is related to the Beltrami differential
as $\mu^z_{\zbar} = \del_{\zbar}v^z$.

Since ${U}$ has no poles with the $b$ ghost
(as is the case in the bosonic string),
use of this canonical prescription is natural.
Moreover, the prescription~(\ref{eq:PS1loopA}) has
the full BRST invariance so,
barring the usual concern with the moduli boundary
contribution, arbitrary BRST trivial pieces
may be added to the vertex operators.
In particular, one can go to the minimal gauge ($Q_{1}=0$)
and there the prescription is well-tested to give
the correct answers.

\bigskip
To convert the unintegrated vertex $V(w_1)$ in~(\ref{eq:PS1loopA})
to an integrated one,
we first average over its position $w_1$:
\begin{align}
\label{eq:PS1loopA2}
\calA_{n}
&= \int{\mathd^{2}\tau\over\Im\tau}
\Bigl(\prod_{i=1}^{n}\int\mathd^{2}w_{i} \Bigr)
  \bigl|\langle\calN_{0}\calN'_{0}
    \oint_A\mathd z b(z) V(w_1)
     \Bigl( \prod_{i=2}^{n} {U}(w_{i})\Bigr)\rangle
  \bigr|^{2} \,.
\end{align}
Note that $\Im\tau$ is the area of the torus of modulus $\tau$.
If we were dealing with the bosonic string,
a zero-mode of $c$ ghost can be split off
from the unintegrated vertex, $V =cU$,
and~(\ref{eq:bosonic1loopB}) is essentially derived.
However, in the pure spinor formalism, there is no
$c$ ghost so we wish to use the $b$ ghost present in~(\ref{eq:PS1loopA2})
to convert $V$ to ${U}=b_{-1}V$.
Therefore we rewrite
\begin{align}
  \oint_{A} \mathd z b(z)
 &= -\oint_{C}\mathd z'\oint_{A}\mathd z b(z')J_{g}(z)
\end{align}
where $C$ is a contour that surrounds $z$.
Then, pulling the contour $C$ off $z$, we get
\begin{align}
 \calA
\label{eq:newprescription1}
&= \int{\mathd^{2}\tau  \over \Im\tau}
 \int\Bigl(\prod_{i=1}^{n}\mathd^{2}w_{i}\Bigr)
\bigl|\langle\calN_{0}\calN'_{0} \oint_{A}\mathd z J_{g}(z)
   \prod_{i=1}^{n}{U}(w_{i})\rangle \bigr|^{2}\,
\end{align}
where we have used that $b$ has no poles with $U$.
This is our prescription for the $n$-point $1$-loop
amplitudes that treats all external vertices equally.

Note that~(\ref{eq:newprescription1}) is valid only for
the vertices in a special class of gauges,
because not all BRST trivial operators decouple anymore.
This explains why the minimal gauge vertices cannot be used
in~(\ref{eq:newprescription1}).
However,~(\ref{eq:newprescription1}) still has
a residual gauge invariance since
operators of the form $Q(b_0\Omega)$ decouple.

The derivation here of course applies to the bosonic string
as well.
There, since $b$ and $c$ in $J_{g}$ contribute
only the zero-modes, the integration of $J_{g}(z)$
over $A$ can be undone,
\begin{align}
  \oint_{A}\mathd zJ_{g}(z) = -bc(y)\,,\quad
\text{($y$: arbitrary point)}
\end{align}
and hence,
\begin{align}
\calA
&= \int{\mathd^{2}\tau  \over \Im\tau}
 \int\Bigl(\prod_{i=1}^{n}\mathd^{2}w_{i}\Bigr)
\bigl|\langle J_{g}(y)
   \prod_{i=1}^{n}U(w_{i})\rangle \bigr|^{2}
\end{align}
as is well known~\cite{Polchinski:1998rq}.
However, we note again that integrated vertices in this formula
are no longer allowed to be in an arbitrary gauge;
they must be related to the conventional representatives
(e.g.\ $U=\mathe^{\mathi k\cdot x}$ for the tachyon)
by a gauge transformation of the type $\delta U = Q(b_0\Omega)$.

\subsection{Residual BRST invariance of the new prescription}

In the derivation of this prescription,
it was important that the integrated vertex ${U}$
was annihilated by $b_{-1}$.
Therefore, it seems that one is no longer free to choose
an arbitrary gauge for vertices by adding BRST trivial pieces.
However, it will now be argued that,
up to a possible contribution from the boundary of the
moduli space, ${U}$ still has a residual gauge invariance of the form
\begin{align}
\delta {U} &= Q(b_0 \Omega)
\end{align}
where $\Omega$ is an arbitrary weight $1$ primary operator.

To show that operators of the form $Q(b_0\Omega)$ decouple
from the amplitude, consider a variation of
an $(n+1)$-point amplitude:
\begin{align}
\delta\calA &= \langle \oint_A J_g\, U_1\cdots U_n\, Q(b_0\Omega) \rangle\,.
\end{align}
Here $U_i\equiv U_i(w_i)$ and we omitted the integrations
over $w_{i}$'s.
Since the $U_{i}$'s are $Q$-closed under the integration symbol,
we treat them as if they are $Q$-closed.

Now, pulling the contour of $Q$ and $b_{0}$ off of $\Omega$
and using that $Q(U_i) = b_0(U_i)=0$,
we find
\begin{align}
\label{eqn:X1}
\delta\calA
&= -\langle \oint_A J_B\, U_1\cdots U_n\, b_0\Omega \rangle\\
&= 
 \langle \oint_A b_0(J_B)\, U_1\cdots U_n\, \Omega \rangle\\
&= 
 \langle \oint_A L_0\, U_1\cdots U_n\, \Omega \rangle\,
\end{align}
where $J_{B}$ denotes the BRST current.
Note that on a torus, $b_0(X)$ can be written as 
$[\oint_A b , X] $  so that $b_0 (X_1 X_2) = b_0(X_1)\, X_2 \pm X_1\,b_0 (X_2)$.
Since all vertex
operators are primary fields, insertion of $\oint_{A}L_{0}$ generates
a total derivative on the moduli space.
so $Q(b_0\Omega)$ indeed decouples from the amplitude.

\subsection{$4$-point $1$-loop massless amplitude}

It will now be shown
that if
one uses the integrated vertex operator $U$
in Siegel gauge.
the new prescription of section~\ref{subsec:prescription}
gives a non-vanishing result for the
$4$-point $1$-loop massless amplitude.
Below, we only write the chiral half of the closed string
and use the terminology appropriate for the open string.

We first show that the only non-vanishing contribution
comes from the four product of the $d_{\alpha}r^{3}$
and $N_{mn}r^{4}$ terms
\begin{align}
d_{\alpha}{r^3D^{2}A_{\alpha\beta} \over (\overline{\lambda}\lambda)^{3}}
+ N_{mn}{r^4D^{2}A_{\alpha\beta} \over (\overline{\lambda}\lambda)^{4}}
\end{align}
in $U$.
Note that these get smeared to
\begin{align}
\label{eqn:smeareddN}
d_{\alpha}{(r+\overline{g})^{3}
 D^{2}A_{\alpha\beta}(x,\theta+ g )\over
|\lambda+f|^{6}}
+N_{mn}{(r+\overline{g})^{4}
 D^{2}A_{\alpha\beta}(x,\theta+ g )\over
|\lambda+f|^{8}}
\end{align}
in the presence of the extra regulator $\calN'(y)$ of
section~\ref{sec:regularization}.
The regularization of $(\overline{\lambda}\lambda)^{-L}$ pole
at the same time shifts
$r^{L}$ to $(r+\overline{g})^{L}$,
so combinations with $L>11$ can give non-zero contribution.

To show that these are the only contributions to the amplitude,
first recall that, for each of the products in $(\int U)^4$,
only one of the two regularization factors
$\calN_{0}$ and $\calN'_{0}$
is necessary.
The former is needed when the total order of $(\overline{\lambda}\lambda)$ poles
(or equivalently, the total $r$-degree)
is below $(\overline{\lambda}\lambda)^{-11}$,
and the latter is needed when it exceeds $(\overline{\lambda}\lambda)^{-11}$.\footnote{
For the eleventh pole $r^{11}/(\overline{\lambda}\lambda)^{11}$
see the discussion at the end of section~\ref{subsec:regtwo}.
However, since this $r^{11}$ term cannot saturate the
fermionic zero modes for the $4$-point $1$-loop amplitudes,
one may ignore this subtlety here.}

Since the regulator
\begin{align}
\label{eqn:calNzzz}
\calN_{0} = \exp[-(\overline{\lambda}\lambda+r\theta)-(\overline{N}N+(\overline{\lambda}\lambda)sd)]
\end{align}
can only provide at most $11$ $d_{\alpha}$ zero modes,
and $4$ $U$'s can provide at most $4$ $d_{\alpha}$ zero modes,
it is clear that combinations of the terms
for which the total order of $(\overline{\lambda}\lambda)$ pole $L$
is below $11$ cannot contribute.
Therefore, we can forget about the
terms requiring $\calN_{0}$ regularization
(and $\calN_{0}$ itself).

For combinations of the terms requiring
the $\calN'$ regularization, 
one has to saturate the fermionic zero modes
of $(d_{\alpha},\theta^{\alpha})$, $(s_{\alpha},r^{\alpha})$
and $(g^{\alpha},\overline{g}_{\alpha})$
to have a non-vanishing result.
Unlike the $\calN$ regulator of~(\ref{eqn:calNzzz}),
the zero mode remnant of the $\calN'$ regulator
\begin{align}
\calN'_{0}
&= \exp[{-(\overline{\omega}\omega+sd)}
 + {(f\omega+gd)}
 + (\overline{f}\overline{\omega}+\overline{g}s)]
\end{align}
can provide more than $11$ $d_{\alpha}$ zero modes
because the zero modes of $d_{\alpha}$ appear
both in $\exp(-sd)$ and $\exp(gd)$.
We will find that $\calN'_{0}$ can provide
$L$ $d_{\alpha}$ zero modes for the combination of
terms that goes as $r^{L}/(\overline{\lambda}\lambda)^{L}$
in the absence of $\calN'$ regularization.

We first show that the $r^{12}$ term
in $(d_{\alpha}f^{\alpha})^{4}$ can saturate all the zero-modes.
In the gauge $(\overline{\lambda}\gamma^{m})^{\alpha}A_{\alpha\beta}=0$,
$f^{\alpha}$ contains terms up to $r^{3}/(\overline{\lambda}\lambda)^{3}$,
so $(d_{\alpha}f^{\alpha})^{4}$ contains terms up to $r^{12}/(\overline{\lambda}\lambda)^{12}$.
However, as explained just above,
terms with total $r$-degree below $11$
cannot saturate $d_{\alpha}$ zero modes,
so we only need to keep 
$r^{11}/(\overline{\lambda}\lambda)^{11}$
and $r^{12}/(\overline{\lambda}\lambda)^{12}$.

The coefficients of $(\overline{\lambda}\lambda)^{-L}$ ($L=11$ or $12$)
in these combinations are
\begin{align}
\label{eqn:poleorderL}
\mathe^{-(\overline{\omega}\omega+sd)}
\mathe^{(f\omega+gd) +(\overline{f}\overline{\omega}+\overline{g}s)}
\times (d)^4(r+\overline{g})^{L}
  D^{20-L}A^{4}_{\alpha\beta}(x,\theta+g)
\end{align}
where the exponential factors come from the regulator
$\calN'_{0}$, and the rest come from
the smeared vertices~(\ref{eqn:smeareddN}).
Now, to saturate
all the zero modes of $r_{\alpha}$, $s^{\alpha}$ and $\overline{g}_{\alpha}$,
one has to take the following combination
in~(\ref{eqn:poleorderL}):
\begin{align}
\label{eqn:saturation}
\begin{split}
&\mathe^{-\overline{\omega}\omega}\mathe^{(f\omega+gd)+\overline{f}\overline{\omega}}
(sd)^{L-11}(\overline{g}s)^{22-L}
\times (d)^{4}(r^{11}\overline{g}^{L-11})D^{20-L}A^{4}_{\alpha\beta}(x,\theta+g) \\
&= \mathe^{-\overline{\omega}\omega}\mathe^{(f\omega+gd)+\overline{f}\overline{\omega}}
d^{L-7} (sr)^{11} \overline{g}^{11}
D^{20-L}A^{4}_{\alpha\beta}(x,\theta+g)  \,.
\end{split}
\end{align}
Then, it is clear that $L$ has to be $12$
in order to saturate the $16$ $d_{\alpha}$ zero modes
and $11$ $g^{\alpha}$ zero modes as in
\begin{align}
\label{eqn:satureted}
\begin{split}
&\mathe^{-\overline{\omega}\omega}\mathe^{f\omega+\overline{f}\overline{\omega}}
(gd)^{11}d^{L-7} (sr)^{11} \overline{g}^{11}
D^{20-L}A^{4}_{\alpha\beta}(x,\theta+g) \\
&= \mathe^{-\overline{\omega}\omega}\mathe^{f\omega+\overline{f}\overline{\omega}}
 d^{L+4} (sr)^{11}(g\overline{g})^{11}
 D^{20-L}A^{4}_{\alpha\beta}(x,\theta) \,.
\end{split}
\end{align}
Thus, in $(d_{\alpha}f^{\alpha})^{4}$, only
\begin{align}
\label{eqn:d4}
\biggl(d_{\alpha}{r^3D^{2}A_{\alpha\beta} \over (\overline{\lambda}\lambda)^{3}}
\biggr)^{4}
\end{align}
contributes to the amplitude.

This counting of the zero modes at the same time
explains that the $\calN'_{0}$ regulator
can provide $L$ $d_{\alpha}$ zero modes for the
term that naively goes like $r^{L}/(\overline{\lambda}\lambda)^{L}$.
Then, it is clear that only
\begin{align}
\label{eqn:dN4}
\biggl(d_{\alpha}{r^3D^{2}A_{\alpha\beta} \over (\overline{\lambda}\lambda)^{3}}
+N_{mn}{r^4D^{2}A_{\alpha\beta} \over (\overline{\lambda}\lambda)^{4}}
\biggr)^{4}
\end{align}
contributes to the $4$-point amplitude
in the $(\overline{\lambda}\gamma^{m})^{\alpha}A_{\alpha\beta}=0$ gauge.
(Other combinations cannot saturate the $16$ $d_{\alpha}$
zero modes, because they are of the form
$d^{k}r^{L-k}/(\overline{\lambda}\lambda)^{L-k}$ with $L<16$
and $k=1,\cdots,4$.)

\bigskip
Let us make a consistency check for
the amplitudes computed using the prescription given here.
The $4$ point amplitude should have the dimension
\begin{align}
\calA \sim F^{4}\,,
\end{align}
where $F=F_{mn}$ is the photon field strengths.
In the superfield $A_{\alpha\beta}(x,\theta)$,
the fieldstrength resides at the $\theta^{6}$ level
so using $A_{\alpha\beta}$, the amplitude should be
\begin{align}
\calA \sim D^{24}A^{4}_{\alpha\beta}
 \sim \int\mathd^{16}\theta D^{8}A^{4}_{\alpha\beta}\,,
\end{align}
and this is what we get by integrating~(\ref{eqn:satureted})
over all zero modes (for $L=12$).

\bigskip
We now show that the amplitude computed
as above is indeed BRST invariant.
To show that one is computing a BRST invariant quantity,
one has to check that the result is invariant under
the BRST variation of the regulator,
\begin{align}
\begin{split}
\calN'(y)
&=
\exp[Q(s^{\alpha}\omega_{\alpha}+g^{\alpha}\omega_{\alpha}+\overline{g}_{\alpha}\overline{\omega}^{\alpha})] \\
\quad\to\quad
\calN'_{c,\varepsilon,\overline{\varepsilon}}(y)
&= \exp[Q(cs^{\alpha}\omega_{\alpha}+\varepsilon g^{\alpha}\omega_{\alpha}
  +\overline{\varepsilon}\overline{g}_{\alpha}\overline{\omega}^{\alpha})) \\
&= \exp[-c(\overline{\omega}_{\alpha}\omega^{\alpha}+s^{\alpha}d_{\alpha}]
  +\varepsilon(f^{\alpha}\omega_{\alpha}+g^{\alpha}d_{\alpha})
  +\overline{\varepsilon}
    (\overline{f}_{\alpha}\overline{\omega}^{\alpha}+\overline{g}_{\alpha}s^{\alpha})] \,,
\end{split}
\end{align}
for some constants $c$, $\varepsilon$, and $\overline{\varepsilon}$.

To check the invariance of~(\ref{eqn:d4})
under this variation of the regulator,
one first notes that
the zero mode products in~(\ref{eqn:saturation})
with $L=12$ scale as
\begin{align}
\begin{split}
&\mathe^{-c(\overline{\omega}\omega+sd)}
\mathe^{\varepsilon(f\omega+gd)
 +\overline{\varepsilon}(\overline{f}\overline{\omega}+\overline{g}s)}
\times (d)^4(r+\overline{\varepsilon}\overline{g})^{12}
  D^{12}A^{4}_{\alpha\beta}(x,\theta+\varepsilon g) \\
&=\mathe^{-c(\overline{\omega}\omega)}
\mathe^{\varepsilon (f\omega)
 +\overline{\varepsilon}(\overline{f}\overline{\omega})}
 (c sd)^{1}
 (\varepsilon gd)^{11}(\overline{\varepsilon}\overline{g}s)^{10}
\times (d)^4(\overline{\varepsilon}^{1} r^{11}\overline{g}^{1})
  D^{12}A^{4}_{\alpha\beta}(x,\theta)
 \\
&= c^{1}(\varepsilon\overline{\varepsilon})^{11}
 \mathe^{-c(\overline{\omega}\omega)}
\mathe^{\varepsilon (f\omega)
 +\overline{\varepsilon}(\overline{f}\overline{\omega})}
  (g\overline{g})^{11}d^{16}
  (s r)^{11}
  D^{12}A^{4}_{\alpha\beta}(x,\theta) \,.
\end{split}
\end{align}
So to show the BRST invariance,
one needs to check that the bosonic integrations
provide $c^{-1}(\varepsilon\overline{\varepsilon})^{-11}$:
\begin{align}
\begin{split}
&\int\mathd^{22}\omega\mathd^{22}\lambda\mathd^{22}f
 \,\mathe^{-c(\overline{\omega}\omega)
 +\varepsilon(f\omega)+\overline{\varepsilon}(\overline{f}\overline{\omega})}
 \prod_{i=1}^{4}{1\over |\lambda+\varepsilon f_{i}|^{6}} 
 \sim c^{-1}(\varepsilon\overline{\varepsilon})^{-11}\,.
\end{split}
\end{align}
This scaling can be easily shown
by performing the change of variables
\begin{align}
(\omega,\lambda,f)\to (\omega',\lambda',f')= (c^{1/2}\omega,c^{-1/2}\lambda,\varepsilon c^{-1/2}f)
\end{align}
so that the integral becomes
\begin{align}
&c^{-1}(\varepsilon\overline{\varepsilon})^{-11}\times
 \int \mathd^{22}\omega'\mathd^{22}\lambda'\mathd^{22}f'
\, \mathe^{-(\overline{\omega}'\omega')
 +(f'\omega')+(\overline{f}'\overline{\omega})'}
 \prod_{i=1}^{4}{1\over |\lambda'+ f'_{i}|^{6}} \,.
\end{align}
Similarly, all contributions from~(\ref{eqn:dN4})
can be checked to be invariant under the BRST variation
of the $\calN'$ regulator.

\bigskip
To summarize, we have shown that
the $4$-point $1$-loop amplitude can be computed
using $4$ integrated vertex operators $U$ in the Siegel gauge.
To be able to do so, it was important that the
$U$'s are conformal primaries of weight $1$
and are annihilated by $b_{-1}$.
This explains why one could not compute
the amplitude using $4$ integrated vertex operators
in the minimal gauge.

Since the order of $(\overline{\lambda}\lambda)$ poles
in $(\int U)^4$ exceeds $11$,
a regularization for $(\overline{\lambda}\lambda)\to0$ was necessary.
We used a regularization method proposed in~\cite{Berkovits:2006vi}
(and explained in section~\ref{sec:regularization})
to define the indefinite factor of the form
\begin{align}
\int\mathd^{22}\lambda\mathd^{11} {r^{L}\over(\overline{\lambda}\lambda)^{L}}\,,\quad
(L>11)\,.
\end{align}
Moreover, since the Siegel gauge integrated vertex operator
takes a relatively simple form in $(\overline{\lambda}\gamma^{m})^{\alpha}A_{\alpha\beta}=0$ gauge,
we were able to identify the combinations of the terms
that contributes to the amplitude
and their invariance under the BRST variation of the regulator $\calN'$.
Our conclusion is that the only contribution comes from
the terms~(\ref{eqn:dN4}) that require $\calN'$ regularization
for $(\overline{\lambda}\lambda)\to0$.

Finally, let us mention that we have demonstrated
(ignoring the gauge invariance $\delta\omega_{\alpha}=(\gamma^{m}\lambda)_{\alpha}\Omega_{m}$)
that the non-zero modes of $s^{\alpha}$ in the regulator $\calN'$
does convert ``extra'' $r_{\alpha}$ zero modes above $11$
to $d_{\alpha}$ zero modes
as was advocated in~\cite{Berkovits:2006vi}.

\section{Summary}
\label{sec:summary}

In this paper, we showed how to construct vertex operators
in the pure spinor formalism in the Siegel gauge.
Unintegrated vertices in the Siegel gauge can be constructed
as $V_{S}=b_{0}V^{\ast}$, where $V^{\ast}$ is the ghost number $2$
vertex of the corresponding antifield.
Integrated vertices can then be constructed as usual
by $\int U_{S} = \int b_{-1}V_{S}$.

The construction is not obstructed by the 
complexity of the $b$-ghost of the formalism
and works for vertices of all mass levels,
provided that the space of pure spinor vertices has
field-antifield doubling.
Although this latter fact is non-trivial in the pure spinor formalism,
it is strongly supported by the study of the partition function
of the pure spinor operator space in~\cite{Berkovits:2005hy,
Grassi:2005gg,Aisaka:2008vw}.

For the massless states, an explicit form of the 
antifield vertex operator is 
$V^{\ast}=\lambda^{\alpha}\lambda^{\beta}A_{\alpha\beta}(x,\theta)$,
and the computation of the Siegel gauge vertices
(both unintegrated and integrated) is straightforward. Although
the form of the integrated vertex operator $U_{S}$
is fairly complicated,
we showed that in the gauge where $A_{\alpha\beta}$
satisfies $(\overline{\lambda}\gamma^{m})^{\alpha}A_{\alpha\beta}=0$,
the form of $U_{S}$ simplifies considerably
and contains terms only up to $r^{4}/(\overline{\lambda}\lambda)^{4}$.

\bigskip
When vertices are in the Siegel gauge,
it is well-known in bosonic string theory that
the $n$-point $1$-loop amplitude can be computed
using $n$ integrated vertex operators.
We have shown that this Siegel gauge prescription is also valid
in the pure spinor
formalism by deriving it
from the conventional prescription that uses
$1$ unintegrated and $(n-1)$ integrated vertex operators.

This new $1$-loop prescription provides
a good testing ground for the regularization prescription
of~\cite{Berkovits:2006vi} (reviewed in section~\ref{sec:regularization})
for the functional integration
region $(\overline{\lambda}\lambda)\sim 0$.
This regularization becomes necessary when
the factor of $r/(\overline{\lambda}\lambda)$
in the integrand accumulates to
$r^{11}/(\overline{\lambda}\lambda)^{11}$
or higher.
Since the Siegel gauge vertex operators 
have poles in $(\overline{\lambda}\lambda)$,
the $4$-point $1$-loop amplitude already requires
this regularization of $(\overline{\lambda}\lambda)\sim 0$.
Although we have not worked out the explicit
index contractions,
we identified the combinations of terms in $4$ $U_{S}$'s
(in the $(\overline{\lambda}\gamma^{m})^{\alpha}A_{\alpha\beta}=0$ gauge)
that can contribute to the amplitude,
and argued that they give a well defined quantity.

Note that if one blindly applies the new $1$-loop prescription
to the ``minimal gauge'' vertices
that do not depend on non-minimal variables,
one would get a vanishing $4$-point amplitude
because of an undersaturation of $d_{\alpha}$ zero modes.
For the Siegel gauge vertices, we observed that
the regularization of $(\overline{\lambda}\lambda)\sim 0$
converts the extra factors of $r_{\alpha}$'s to
$d_{\alpha}$ zero modes, and the
correct saturation of all fermionic zero modes
is realized.

\bigskip
There are several possible continuations
of the present work.
Firstly, it should be possible to complete the
computation of the $4$-point $1$-loop amplitude
using the new prescription by working out
the index contractions explicitly. Although the
regularization prescription of~\cite{Berkovits:2006vi}
becomes more complicated when one makes it consistent
with the pure spinor constraint,
the number of terms that contributes to the
amplitude should not change and is fairly small.

Secondly, since we now have a method to construct
Siegel gauge vertex operators systematically,
it might be possible to obtain a gauge fixed action of
the cubic open superstring field theory proposed in~\cite{Berkovits:2005bt}.

\bigskip
{\bfseries Acknowledgements: } We would like to thank
Pietro Antonio Grassi, Joost Hoogeveen,
Carlos Roberto Mafra, Nikita Nekrasov, Warren Siegel
and Pierre Vanhove for useful conversations, and the KITP where
part of this research was done.
YA would like to thank FAPESP grant 06/59970-5 for financial support, and
NB would like to thank CNPq grant 300256/94-9 and FAPESP grant
04/11426-0 for partial financial support. 
This research was supported in part by
the National Science Foundation under Grant No. PHY05-51164.


\end{document}